\documentclass[twocolumn]{aastex63}

\submitjournal{ApJ}

\shorttitle{Dust continuum of $z\sim6$ LBGs}
\shortauthors{Mitsuhashi et al.}

\graphicspath{{./}}
\renewcommand\baselinestretch{0.92}
\usepackage{amsmath}
\tabletypesize{\footnotesize}
\usepackage{multirow}

\usepackage{ulem} 
\def\blue#1 {{\textcolor{blue}{#1}}\ }
\def\cii{[C\,{\sc ii}]}
\def\oiii{[O\,{\sc iii}]}
\def\kms{\,km\,s$^{-1}$}

\def\red#1 {{\textcolor{red}{#1}}\ }

\begin{document}

\title{SERENADE II: An ALMA Multi-Band Dust-Continuum Analysis of 28 Galaxies at $5<z<8$ and\\ the Physical Origin of the Dust Temperature Evolution}

\correspondingauthor{Ikki Mitsuhashi}
\email{ikki0913astr@gmail.com}

\author{Ikki Mitsuhashi}
\affiliation{Department of Astronomy, The University of Tokyo, 7-3-1 Hongo, Bunkyo, Tokyo 113-0033, Japan}
\affiliation{National Astronomical Observatory of Japan, 2-21-1 Osawa, Mitaka, Tokyo 181-8588, Japan}

\author{Yuichi Harikane}
\affiliation{Institute for Cosmic Ray Research, The University of Tokyo, 5-1-5 Kashiwanoha, Kashiwa, Chiba 277-8582, Japan}

\author{Franz E. Bauer}
\affiliation{Instituto de Astrof{\'{\i}}sica and Centro de Astroingenier{\'{\i}}a, Facultad de F{\'{i}}sica, Pontificia Universidad Cat{\'{o}}lica de Chile, Campus San Joaquín, Av. Vicuña Mackenna 4860, Macul Santiago, Chile, 7820436}
\affiliation{Millennium Institute of Astrophysics, Nuncio Monse{\~{n}}or S{\'{o}}tero Sanz 100, Of 104, Providencia, Santiago, Chile}
\affiliation{Space Science Institute, 4750 Walnut Street, Suite 205, Boulder, Colorado 80301}

\author{Tom Bakx}
\affiliation{Department of Space, Earth and Environment, Chalmers University of Technology, Gothenburg, Sweden}

\author{Andrea Ferrara}
\affiliation{Scuola Normale Superiore, Piazza dei Cavalieri 7, I-56126, Pisa, Italy}

\author{Seiji Fujimoto}
\affiliation{Department of Astronomy, The University of Texas at Austin, 2515 Speedway Boulevard Stop C1400, Austin, TX 78712-1205, USA}

\author{Takuya Hashimoto}
\affiliation{Tomonaga Center for the History of the Universe (TCHoU), Faculty of Pure and Applied Sciences, University of Tsukuba, Tsukuba, Ibaraki 305-8571, Japan}

\author{Akio K. Inoue}
\affiliation{Waseda Research Institute for Science and Engineering, Faculty of Science and Engineering, Waseda University, 3-4-1, Okubo, Shinjuku, Tokyo 169-8555, Japan}
\affiliation{Department of Physics, School of Advanced Science and Engineering, Faculty of Science and Engineering, Waseda University, 3-4-1, Okubo, Shinjuku, Tokyo 169-8555, Japan}

\author{Kazushi Iwasawa}
\affiliation{Institut de Ci{\`e}ncies del Cosmos (ICCUB), Universitat de Barcelona (IEEC-UB)} \affiliation{Mart{\'i}i Franqu{\`e}s, 1, 08028 Barcelona, Spain}
\affiliation{ICREA, Pg. Llu{\'i}s Companys 23, 08010 Barcelona, Spain}

\author{Yuri Nishimura}
\affiliation{Department of Astronomy, The University of Tokyo, 7-3-1 Hongo, Bunkyo, Tokyo 113-0033, Japan}

\author{Masatoshi Imanishi}
\affiliation{National Astronomical Observatory of Japan, 2-21-1 Osawa, Mitaka, Tokyo 181-8588, Japan}

\author{Yoshiaki Ono}
\affiliation{Institute for Cosmic Ray Research, The University of Tokyo, 5-1-5 Kashiwanoha, Kashiwa, Chiba 277-8582, Japan}

\author{Toshiki Saito}
\affiliation{National Astronomical Observatory of Japan, 2-21-1 Osawa, Mitaka, Tokyo 181-8588, Japan}

\author{Yuma Sugahara}
\affiliation{National Astronomical Observatory of Japan, 2-21-1 Osawa, Mitaka, Tokyo 181-8588, Japan}
\affiliation{Waseda Research Institute for Science and Engineering, Faculty of Science and Engineering, Waseda University, 3-4-1, Okubo, Shinjuku, Tokyo 169-8555, Japan}

\author{Hideki Umehata}
\affiliation{Institute for Advanced Research, Nagoya University, Furocho, Chikusa, Nagoya 464-8602, Japan}
\affiliation{Department of Physics, Graduate School of Science, Nagoya University, Furocho, Chikusa, Nagoya 464-8602, Japan}
\affiliation{Cahill Center for Astronomy and Astrophysics, California Institute of Technology, MS 249-17, Pasadena, CA 91125, USA}

\author{Livia Vallini}
\affiliation{INAF-Osservatorio di Astrofisica e Scienza dello Spazio, via Gobetti 93/3, I-40129, Bologna, Italy}

\author{Tao Wang}
\affiliation{School of Astronomy and Space Science, Nanjing University, Nanjing, Jiangsu 210093, China}
\affiliation{Key Laboratory of Modern Astronomy and Astrophysics, Nanjing University, Ministry of Education, Nanjing 210093, China}

\begin{abstract}
We present an analysis of ALMA multi-band dust-continuum observations for 28 spectroscopically-confirmed bright Lyman-break galaxies at $5<z<8$. Our sample consists of 11 galaxies at $z\sim6$ newly observed in our ALMA program, which substantially increases the number of $5<z<8$ galaxies with both rest-frame 88 and 158 $\mu{\rm m}$ continuum observations, allowing us to simultaneously measure the IR luminosity and dust temperature for a statistical sample of $z\gtrsim5$ galaxies for the first time. We derive the relationship between the UV slope ($\beta_{\rm UV}$) and infrared excess (IRX) for the $z\sim6$ galaxies, and find a shallower IRX-$\beta_{\rm UV}$ relation compared to the previous results at $z\sim2$--4. 
Based on the IRX-$\beta_{\rm UV}$ relation consistent with our results and the $\beta_{\rm UV}$-$M_{\rm UV}$ relation including fainter galaxies in the literature, we find a limited contribution of the dust-obscured star formation to the total SFR density, $\sim30\%$ at $z\sim6$.
Our measurements of the dust temperature at $z\sim6-7$, $T_{\rm dust}=40.9_{-9.1}^{+10.0}\,{\rm K}$ on average, supports a gentle increase of $T_{\rm dust}$ from $z=0$ to $z\sim6$--7. Using an analytic model with parameters consistent with recent {\it{JWST}} results, we discuss that the observed redshift evolution of the dust temperature can be reproduced by an $\sim0.6\,{\rm dex}$ increase in the gas depletion timescale and $\sim0.4\,{\rm dex}$ decrease of the metallicity. The variety of $T_{\rm dust}$ observed at high redshifts can also be naturally explained by scatters around the star-formation main sequence and average mass-metallicity relation, including an extremely high dust temperature of $T_{\rm dust}>80\,{\rm K}$ observed in a galaxy at $z=8.3$.
\end{abstract}


\section{Introduction}
The cosmic star formation rate density (SFRD) is one of the most important quantities to understand the evolution of galaxies as well as the history of the universe.
Over recent decades, the SFRD has been derived through the UV emission of galaxies from the present day ($z=0$) to $z\sim12$ by utilizing ground-based and space telescopes such as the Canada-France-Hawaii Telescope (CFHT), Keck Telescope, Subaru Telescope, {\it Hubble Space Telescope} ({\it HST}) and {\it James-Webb Space Telescope} \citep[{\it JWST}; e.g.,][]{1996ApJ...460L...1L,1996MNRAS.283.1388M,1999ApJ...519....1S,2007ApJ...670..928B,2010ApJ...725L.150O,2013ApJ...773...75O,2013ApJ...762...32C,2013ApJ...763L...7E,2015ApJ...803...34B,2015ApJ...810...71F,2020MNRAS.494.1894M,2023MNRAS.523.1009B,2023MNRAS.523.1036B,2023arXiv230406658H,2023ApJS..265....5H}.
These observations revealed that the SFRD traced by rest-frame UV emission peaks at $z\sim2$--3 and subsequently declines toward the early Universe \citep[][and references therein]{2014ARA&A..52..415M}.

Given that the UV radiation from young massive stars within galaxies is easily absorbed by interstellar dust, it is necessary to measure both direct and re-processed stellar emission corresponding to both the rest-frame ultraviolet (UV) and the rest-frame far infrared (FIR), respectively.
The dust-obscured SFRD is explored with FIR to sub-mm observatories such as {\it Spitzer}, {\it Herschel}, and the James Clark Maxwell Telescope (JCMT), revealing a monotonic increase by an order of magnitude from $z=0$ to $z\sim2$ \citep[e.g.,][]{2001ApJ...556..562C,2009A&A...496...57M,2011A&A...528A..35M,2013A&A...553A.132M}.
However, at $z>4$, the dust-obscured SFRD is poorly constrained, with the discrepancy among studies reaching nearly two orders of magnitude \citep[][]{2010Natur.468...49R,2020A&A...643A...8G,2021A&A...649A.152K,2021ApJ...909..165Z,2021ApJ...923..215C,2021Natur.597..489F,2023MNRAS.518.6142A,2023MNRAS.522.3926B,2023arXiv230301658F}.
The relation between infrared excess (IRX$\equiv L_{\rm IR}/L_{\rm UV}$) and the UV spectral slope ($\beta_{\rm UV}$) can be used to correct for the dust-absorbed UV emission in the absence of the FIR/sub-mm observations. 
Since this relationship is useful for inferring the total (UV+IR) star formation activity of galaxies from the rest-frame UV observations alone, it has been extensively explored using several types of local galaxies such as local starbursts \citep{1999ApJ...521...64M,2000ApJ...533..682C,2012ApJ...755..144T} and Small Magellanic Cloud \citep[SMC;][]{2003ApJ...594..279G}, and galaxies at $z\sim1$--3 \citep{2015ApJ...806..259R,2016A&A...587A.122A,2018ApJ...853...56R,2019A&A...630A.153A}.
With the advent of the Atacama Large Millimeter-submillimeter Array (ALMA), some studies investigated the IRX-$\beta_{\rm UV}$ relation at $z>4$.
For example, \citet{2020A&A...643A...4F} examined the IRX-$\beta_{\rm UV}$ relation at $z\sim5$, and found a redshift evolution at $z>4$.
Recent ALMA observations towards $z\sim7$ UV-selected galaxies suggest a significant contribution of the dust-obscured star formation at high redshift \citep{2021Natur.597..489F,2023MNRAS.518.6142A,2023MNRAS.522.3926B}.

While these ALMA studies shed light on the dust-obscured star formation at $z>4$, the IR luminosity depends on their assumption of the shape of FIR spectral energy distribution (SED).
The variations in FIR SEDs are challenging to account for at $z>4$ due to sensitivity limitations in observing the peak of dust thermal emission at mid-to-far IR wavelengths in the observed frame. 
As the FIR SED might depend on the UV slope $\beta_{\rm UV}$ \citep[e.g.,][]{2019A&A...630A.153A}, considering its variation is essential for constructing an accurate IRX-$\beta_{\rm UV}$ relationship.
Specifically, the luminosity-weighted dust temperature ($T_{\rm dust}$) is one of the key parameters that determine the shape of the FIR SEDs.
For instance, the dust temperature severely affects estimates of $L_{\rm IR}$ and therefore the dust-obscured SFR.
Moreover, in the case of the UV-to-FIR SED modeling based on the energy balance principle with a single ALMA observation, the assumed $T_{\rm dust}$ can influence inferred stellar ages or metallicities due to the degeneracy with dust attenuation. 
While there is observational \citep[e.g.,][]{2014A&A...561A..86M,2015A&A...573A.113B,2018A&A...609A..30S,2020MNRAS.498.4192F,2021MNRAS.503.4878S,2022MNRAS.513.3122S,2022MNRAS.516L..30V} and theoretical \citep[e.g.,][]{2017MNRAS.471.5018F,2019MNRAS.487.1844M,2019MNRAS.489.1397L,2020MNRAS.497..956S} evidence suggesting higher $T_{\rm dust}$ in higher-$z$ galaxies, the redshift dependence at $z\gtrsim5$ is still unclear because of the limited sample size \citep{2020MNRAS.498.4192F,2022ApJ...934...64A,2022MNRAS.515.1751W,2023MNRAS.518.6142A}.
Theoretical studies suggest that $T_{\rm dust}$ could be related to galaxies' metal content or star formation surface densities \citep{2019MNRAS.487.1844M,2019MNRAS.489.1397L}, although this has not been confirmed observationally.
Hence, obtaining representative $T_{\rm dust}$ values and examining the potential origin of its variation is essential as a prior assumption to calculate SFR and $L_{\rm IR}$.

In this paper, we examine the FIR property of galaxies at $z\sim6$ with our new ALMA Band-6/8 observations toward 11 Lyman-Break Galaxies (LBGs) at $z_{\rm spec}\sim6$.
The ALMA Band-8, covering rest-frame $\sim90\mu{\rm m}$ dust continuum near the peak of the dust thermal emission at $z\sim6$, enables us to constrain $T_{\rm dust}$.
In addition to individual measurements, stacking analysis can derive an IRX-$\beta_{\rm UV}$ relationship with the consideration of the variety of the FIR SEDs at $z\sim6$.
We aim to inform the IRX-beta relation at $z\sim6$ by comparing our multi-band analysis with single-band measurements in the literature.
We also examine the representative $T_{\rm dust}$ with a statistically significant number of $z\sim5$--8 galaxies by compiling our new observations and archival data of additional 14 galaxies at $z=5$--8.
We re-analyze the archival data in a homogeneous way to our target galaxies at $z\sim6$ for a fair comparison.
An observational attempt to constrain $T_{\rm dust}$ and their evolution yields fundamental information to understand the dust-obscured star formations in the early Universe.

The paper is organized as follows: Section 2 provides an overview of the datasets used in this work. Section 3 describes the method of measurements for UV and IR properties, and shows the parameter coverage of this work compared with previous studies. In Section 4, we report the IRX-$\beta_{\rm UV}$ and $M_{\rm UV}$-$f_{\rm obs}$ relations, the contribution of the dust-obscured star formation at $z\sim6$, and the results of $T_{\rm dust}$ measurements. Discussions of the model of $T_{\rm dust}$ evolution are described in Section 5. The conclusions are presented in Section 6. Throughout this paper, we assume a flat universe with the cosmological parameters of $\Omega_{\rm M}=0.3$, $\Omega_{\Lambda}=0.7$, $\sigma_{8}=0.8$, and $H_0=70$ \kms ${\rm Mpc}^{-1}$.

%
%
%
%
%
%
\section{Observation and Data}
\subsection{SERENADE overview}
SERENADE (Systematic Exploration in the Reionization Epoch using Nebular And Dust Emission) is an ALMA program (ID:\#2022.1.00522.S, PI: Yuichi Harikane) designed to observe the two brightest FIR fine structure lines (\cii\ 158$\mu$m and \oiii\ 88$\mu$m) from LBGs at $z\sim6$.
The parent sample of the target galaxies is selected from the literature, mostly from the galaxy sample identified in Hyper Suprime-Cam Subaru Strategic Program \citep[HSC-SSP;][]{2016ApJ...828...26M,2018PASJ...70S...4A,2018PASJ...70S..35M,2018ApJS..237....5M,2018PASJ...70S..10O,2022ApJS..259...20H}, based on the following four criteria: (1) spectroscopic confirmation of their redshifts, (2) redshifted \oiii\ 88$\mu$m and Ly$\alpha$ falling into ALMA Band-8 and ground-based optical telescopes, respectively ($z=5.8$--6.3), (3) the absence of clear AGN/QSO signatures in their rest-frame UV spectra, and (4) sufficient expected brightness for the detection of emission lines within a relatively short observing time ($\lesssim10\,$hours).
We select 19 galaxies as our targets to encompass a broad dynamic range in a Star Formation Rate (SFR), including two lensed galaxies (see Table \ref{tab:tab1}).
Observations were conducted between October and December 2022 using the C43-1/2/3 configuration.
Since one targeting galaxy (J023536-031737) was not observed in either Band-6 or Band-8, we address the remaining 18 LBGs in this paper.
Among 18 LBGs, 11 are observed in both 158\,$\mu$m and 88\,$\mu$m.

We have also incorporated three galaxies reported in \citet{2020ApJ...896...93H} for the $z\sim6$ galaxy sample with the 158\,$\mu$m and 88\,$\mu$m coverage. 
In total, the SERENADE galaxies are composed of 21 galaxies (18 galaxies in Cycle-9 and 3 galaxies in the pilot study in \citealt{2020ApJ...896...93H}) at $z\sim6$, including 14 galaxies (11 from Cycle-9 and 3 from \citealt{2020ApJ...896...93H}) with the 158\,$\mu$m and 88\,$\mu$m coverage.
We provide more details on the continuum detection and flux measurements in Section \ref{subsec:dustflux}.
Among the 21 target galaxies, observations were conducted for 20 \cii\ and 15 \oiii\ emission lines. 
Of these, 16 \cii\ and 10 \oiii\ emission lines were detected with a signal-to-noise ratio (S/N) greater than 3.
A comprehensive discussion of the emission line data is beyond the scope of this paper and will be provided in an upcoming paper (Paper I; Harikane et al. in prep).

\subsection{Addtional $5<z<8$ samples}
We incorporate an additional 14 galaxies at $5<z<8$ with multi-band ALMA coverage in this study; four galaxies at $z\sim5$ \citep{2020MNRAS.498.4192F}, and ten galaxies at $z\sim7$ \citep{2015Natur.519..327W,2017MNRAS.466..138K,2019PASJ...71...71H,2020MNRAS.495.1577I,2021ApJ...923....5S,2021MNRAS.508L..58B,2022MNRAS.515.1751W,2022ApJ...934...64A,2023MNRAS.518.6142A}.
We collect galaxies at $5<z<8$ with both Band-6 and Band-8 observations publicly available in the ALMA science archive.
For a fair comparison, we re-analyze these datasets in the same manner as the SERENADE galaxies.
The differences between our measurement and previous studies mainly come from the flux measurement methodology, but we note that our results are broadly consistent with the previously mentioned studies.

\subsection{Summary of the samples}
In total, we have 28 galaxies with multi-band ALMA coverage composed of 14 SERENADE galaxies and the additional 14 galaxies at $5<z<8$ in literature.
Among 28 galaxies, 21 are individually detected, and $T_{\rm dust}$ can be estimated.
This is the largest sample of $z\gtrsim5$ galaxies with multi-band dust continuum observation.

%
%
%
%
%
%
\section{Analysis}\label{sec:analysis}
\subsection{Measurement of $\beta_{\rm UV}$ and $M_{\rm UV}$}\label{subsec:betauv}
To compare the dust continuum properties with the rest-frame UV continuum properties, we estimate parameters that characterize the UV continuum, such as UV continuum slope ($\beta_{\rm UV}$; $f_{\lambda}\propto\lambda^{\beta_{\rm UV}}$) and absolute magnitude ($M_{\rm UV}$).
The slope of the UV spectrum is mainly determined by the stellar age, stellar metallicity, and the amount of dust attenuation. 
Our calculations utilize broad-band photometry spanning a rest-frame wavelength range as large as 800--3000\AA\ depending on the data availability at the position of each galaxy. 
All of the utilized photometries, along with their references, are listed in Table \ref{tab:tab1}.
We cross-match catalog coordinates with the central coordinate of our target galaxies derived from the detection in previous papers.
For galaxies residing in the Cosmic Evolution Survey \citep[COSMOS][]{2007ApJS..172....1S}, Subaru/XMM-Newton Deep Field \citep[SXDF][]{2004JCAP...09..011P}, and XMM Large Scale Structure survey \citep[XMM-LSS][]{2004JCAP...09..011P}, we obtain $i$,$z$,$J$,$H$, and $K$-band photometries from the latest COSMOS2020 catalog \citep{2022ApJS..258...11W}, SPLASH-SXDF multi-wavelength catalog \citep{2018ApJS..235...36M}, and VISTA Deep Extragalactic Observations catalog \citep{2013MNRAS.428.1281J}, respectively.
We take $i$,$z$, and $y$-band photometry from the HSC-PDR3 catalog \citep{2022PASJ...74..247A} for the galaxies originally identified in the wide layer of the HSC-SSP survey \citep{2018PASJ...70S...4A}.
If the galaxies are covered in the United Kingdom Infrared Telescope Deep Sky Survey \citep[UKIDSS,][]{2007MNRAS.379.1599L} or VISTA Kilo-degree Infrared Galaxy survey \citep[VIKING,][]{2013MNRAS.428.1281J}, we combine the photometries at NIR wavelengths with the $i$,$z$, and $y$-band photometry following \citet{2019ApJ...883..183M}.
For galaxies not covered by the aforementioned catalogs, we refer to the photometries measured in the literature \citep[mostly from {\it HST} observations, e.g.,][]{2018Natur.553..178S}, or directly adopt $\beta_{\rm UV}$ and/or $M_{\rm UV}$ values \citep[e.g.,][]{2016ApJ...817...11H} when the photometries are not publicly available.


\renewcommand{\arraystretch}{1.2}
\tabcolsep = 0.07cm
%
%
%
%
%
%
\begin{deluxetable*}{cccccccccccc}
\tablecaption{Summary of the parameters related to rest-frame UV spectra \label{tab:tab1}}
\tablewidth{0pt}
\tablehead{\colhead{ID} & 
\colhead{$z_{\rm spec}$\textsuperscript{\textparagraph}} & 
\colhead{$i$\textsuperscript{\textdagger}} & 
\colhead{$z$\textsuperscript{\textdagger}} & 
\colhead{$y$,$Y$,$Y_{110}$\textsuperscript{\textdagger}} & 
\colhead{$J$,$J_{125}$\textsuperscript{\textdagger}} &
\colhead{$H$,$H_{160}$\textsuperscript{\textdagger}} & 
\colhead{$K_{s}$} &
\colhead{$EW^0_{{\rm Ly}\alpha}$} &
\colhead{$M_{1500}$} &
\colhead{$\beta_{\rm UV}$} &
\colhead{ref}\\
 &  & ${\rm [mag]}$ & ${\rm [mag]}$ & ${\rm [mag]}$ & ${\rm [mag]}$ & ${\rm [mag]}$ & ${\rm [mag]}$ & [\AA] & ${\rm [mag]}$ &
}
\startdata
\multicolumn{12}{c}{{\bf SERENADE (Cycle-9)}} \\
J020038-021052 & 6.1120 & $\geq26.7$ & $23.69\pm0.06$ & $\geq25.2$ & - & - & - & $542.23$ & $\geq-21.5$ & $\leq2.03$ & 1,2\\
J021033-052304 & 5.9007 & $25.85\pm0.17$ & $23.78\pm0.06$ & $23.54\pm0.11$ & - & - & - & - & $-23.11\pm0.07$ & $0.25_{-1.05}^{+1.29}$ & 1,3,4\\
J021041-055917 & 5.8216 & $26.61\pm0.27$ & $24.25\pm0.27$ & $24.10\pm0.16$ & - & - & - & - & $-22.54\pm0.13$ & $-0.64_{-2.91}^{+3.15}$ & 1,3\\
J021244-015824 & 6.00 & $25.53\pm0.17$ & $23.23\pm0.03$ & $22.96\pm0.06$ & - & - & - & - & $-23.72\pm0.04$ & $0.33_{-0.48}^{+0.65}$ & 1,4\\
J021735-051032 & 6.12 & - & $25.16\pm0.11$ & $25.40\pm0.26$ & $25.68\pm0.19$ & $26.03\pm0.46$ & $25.81\pm0.28$ & $46.8\pm8.4$ & $-21.23\pm0.08$ & $-2.49_{-0.24}^{+0.24}$ & 6,7\\
J021807-045841 & 6.0446 & $28.04\pm0.61$ & $25.10\pm0.07$ & $25.07\pm0.14$ & $25.54\pm0.13$ & $25.62\pm0.24$ & $26.48\pm0.38$ & $27.6\pm4.1$ & $-21.43\pm0.05$ & $-2.9_{-0.24}^{+0.16}$ & 6,7\\
J021838-050943 & 6.1860 & - & $25.08\pm0.17$ & $24.74\pm0.25$ & $25.02\pm0.15$ & $24.93\pm0.22$ & $25.11\pm0.21$ & $39.4\pm6.0$ & $-21.71\pm0.10$ & $-2.01_{-0.16}^{+0.24}$ & 6,7\\
J022023-050431 & 5.8403 & - & - & $24.22\pm0.04$ & $24.29\pm0.08$ & $24.33\pm0.12$ & $24.29\pm0.19$ & - & $-22.42\pm0.03$ & $-2.17_{-0.08}^{+0.16}$ & 8,9\\
J022627-045238 & 6.0675 & - & $24.77\pm0.04$ & $24.42\pm0.04$ & $24.78\pm0.10$ & $24.28\pm0.09$ & $25.06\pm0.27$ & $13.0\pm4.0$ & $-22.10\pm0.01$ & $-1.61_{-0.08}^{+0.08}$ & 9,10\\
J024801-033258\textsuperscript{\textdaggerdbl}  & 6.0271 & $26.4\pm0.1$ & $25.7\pm0.1$ & $25.30\pm0.09$ & $25.5\pm0.1$ & $25.2\pm0.1$ & - & $21.1\pm2.8$ & $-20.76\pm0.05$ & $-1.44_{-0.16}^{+0.16}$ & 11,12\\
J045408-030014\textsuperscript{\textdaggerdbl} & 6.3163 & - & - & - & - & - & - & $8.2\pm1.4$ & $-20.9\pm0.1$ & $-1.6\pm0.2$ & 13\\
J085723+014254 & 5.8394 & $26.24\pm0.26$ & $24.14\pm0.05$ & $23.93\pm0.09$ & - & - & - & $5.1\pm0.1$ & $-22.71\pm0.07$ & $0.01_{-0.89}^{+0.97}$ & 1,3\\
J091436+044231 & 5.8433 & $25.46\pm0.12$ & $23.29\pm0.06$ & $23.07\pm0.08$ & - & - & - & - & $-23.57\pm0.04$ & $0.09_{-0.89}^{+0.97}$ & 1,14\\
J100008+021136 & 5.8650 & $26.29\pm0.05$ & $24.91\pm0.03$ & $24.78\pm0.04$ & $24.86\pm0.05$ & $24.84\pm0.07$ & $25.00\pm0.10$ & $31.5\pm27.0$ & $-21.79\pm0.01$ & $-2.01_{-0.01}^{+0.01}$ & 15,16\\
J100634+030005 & 5.8588 & $25.79\pm0.10$ & $23.70\pm0.02$ & $23.65\pm0.05$ & - & - & - & - & $-22.99\pm0.03$ & $-1.61_{-0.40}^{+0.57}$ & 1,14\\
J114412-000613 & 6.05 & $\geq26.1$ & $24.63\pm0.10$ & $24.32\pm0.15$ & - & - & - & - & $-22.38\pm0.12$ & $0.25_{-1.70}^{+1.86}$ & 1,8\\
J135348-001026 & 6.1702 & $27.25\pm0.44$ & $23.33\pm0.03$ & $22.65\pm0.03$ & $22.15\pm0.26$ & - & $20.78\pm0.16$ & - & $-24.03\pm0.03$ & $0.49_{-0.16}^{+0.08}$ & 1,17\\
J142824+015934 & 5.9881 & $26.05\pm0.43$ & $22.90\pm0.05$ & $22.83\pm0.07$ & - & - & - & - & $-23.85\pm0.05$ & $-1.61_{-0.73}^{+0.89}$ & 1,14\\
\hline
\multicolumn{12}{c}{{\bf SERENADE pilot} \citep{2020ApJ...896...93H}} \\
J1211+0118 & 6.0293 & $27.82\pm0.76$ & $23.95\pm0.05$ & $23.92\pm0.09$ & - & - & - & $6.9\pm0.8$ & $-22.80\pm0.10$ & $-1.61_{-1.13}^{+0.81}$ & 1,14\\
J0217+0208 & 6.2037 & $\geq26.7$ & $23.90\pm0.04$ & $23.61\pm0.08$ & - & - & - & $15.0\pm1.0$ & $-23.12\pm0.05$ & $-1.20_{-0.81}^{+1.05}$ & 1,14\\
J0235+0532 & 6.0901 & $27.40\pm0.65$ & $23.81\pm0.06$ & $23.89\pm0.14$ & - & - & - & $41.0\pm2.0$ & $-22.80\pm0.12$ & $-0.56_{-1.21}^{+1.45}$ & 1,14\\
\hline
\multicolumn{12}{c}{$z\sim7$ {\bf LBGs} \citep{2019PASJ...71...71H,2021MNRAS.508L..58B,2022ApJ...934...64A,2022MNRAS.515.1751W,2023MNRAS.518.6142A}}  \\
COS-2987030247 & 6.8076 & $27.61\pm0.22$ & $28.46\pm0.81$ & $25.08\pm0.07$ & $24.99\pm0.05$ & $24.81\pm0.06$ & $24.60\pm0.08$ & - & $-21.92\pm0.05$ & $-1.36_{-0.08}^{+0.16}$ & 16,18\\
COS-3018555981 & 6.854 & $\geq27.6$ & - & $25.65\pm0.11$ & $24.98\pm0.04$ & $25.09\pm0.07$ & $25.24\pm0.12$ & - & $-21.91\pm0.02$ & $-2.41_{-0.16}^{+0.16}$ & 16,18\\
UVISTA-Z-001 & 7.0611 & $27.44\pm0.21$ & - & $25.10\pm0.08$ & $23.98\pm0.02$ & $24.02\pm0.03$ & $23.83\pm0.04$ & - & $-22.96\pm0.01$ & $-1.85_{-0.01}^{+0.01}$ & 16,19\\
UVISTA-Z-007 & 6.7498 & $28.20\pm0.40$ & $28.55\pm0.95$ & $24.75\pm0.06$ & $24.74\pm0.04$ & $24.76\pm0.06$ & $24.61\pm0.08$ & - & $-22.13\pm0.02$ & $-1.85_{-0.08}^{+0.08}$ & 16,19\\
UVISTA-Z-019 & 6.7544 & $29.03\pm0.79$ & - & $25.72\pm0.13$ & $25.07\pm0.05$ & $25.21\pm0.08$ & $25.09\pm0.12$ & - & $-21.80\pm0.06$ & $-2.17_{-0.16}^{+0.16}$ & 16,19\\
A1689-zD1 & 7.133 & - & - & $25.00\pm0.13$ & $24.64\pm0.05$ & $24.51\pm0.11$ & - & - & $-22.34\pm0.02$ & $0.01_{-0.73}^{+0.73}$ & 20,21,22\\
B14-65666 & 7.1521 & - & - & - & $24.7\pm0.2$ & $24.6\pm0.3$ & - & - & $-22.29\pm0.18$ & $-1.69_{-1.21}^{+1.21}$ & 23,24\\
REBELS-12 & 7.3459 & - & $26.03\pm0.12$ & $24.92\pm0.06$ & $24.21\pm0.06$ & $24.74\pm0.14$ & $24.14\pm0.11$ & - & $-22.76\pm0.02$ & $-2.09_{-0.16}^{+0.24}$ & 25,26\\
REBELS-25 & 7.3065 & - & $29.08\pm1.63$ & $27.23\pm0.59$ & $25.44\pm0.08$ & $25.22\pm0.10$ & $24.45\pm0.18$ & - & $-21.53\pm0.05$ & $-0.64_{-0.24}^{+0.24}$ & 25,26\\
REBELS-38 & 6.5770 & $\geq27.6$ & $26.88\pm0.23$ & $25.00\pm0.09$ & $25.00\pm0.17$ & $24.84\pm0.20$ & $25.00\pm0.16$ & - & $-21.87\pm0.03$ & $-2.01_{-0.32}^{+0.32}$ & 25,26\\
\hline
\multicolumn{12}{c}{$z\sim5.5$ {\bf LBGs} \citep{2020ApJS..247...61F}} \\
HZ4 & 5.544 & $24.77\pm0.02$ & $24.03\pm0.02$ & $24.05\pm0.03$ & $24.24\pm0.11$ & $23.81\pm0.10$ & $24.32\pm0.10$ & - & $-22.53\pm0.01$ & $-2.09_{-0.08}^{+0.01}$ & 16,27,28\\
HZ6 & 5.293 & $24.40\pm0.02$ & $23.62\pm0.02$ & $23.58\pm0.03$ & $23.47\pm0.02$ & $23.35\pm0.02$ & $23.26\pm0.03$ & - & $-22.90\pm0.01$ & $-1.61_{-0.01}^{+0.01}$ & 16,27,28\\
HZ9 & 5.541 & $25.67\pm0.03$ & $24.83\pm0.03$ & $24.93\pm0.05$ & $24.21\pm0.07$ & $24.50\pm0.12$ & $24.51\pm0.09$ & - & $-21.80\pm0.01$ & $-1.44_{-0.08}^{+0.08}$ & 16,27,28\\
HZ10 & 5.657 & $25.68\pm0.04$ & $24.50\pm0.03$ & $24.38\pm0.04$ & - & - & - & - & $-22.20\pm0.01$ & $-0.88_{-0.40}^{+0.57}$ & 16,27,28\\
\enddata
\tablecomments{
\vspace{-6pt}\tablenotetext{\small \P}{The difference in the number of decimal places reflects the precision. For instance, the redshifts of the SERENADE galaxies determined by \cii/\oiii\ emission lines and Lya emission/UV absorption lines have four and two decimals, respectively.}
\vspace{-6pt}\tablenotetext{\small \dagger}{The $izy$, $Y_{110}J_{125}H_{160}$, and $YHKs$ represent the Subaru/HSC, {\it HST}, and VISTA photometric bands, respectively.}
\vspace{-6pt}\tablenotetext{\small \ddagger}{Lensed galaxies. Lens magnification factors are 2.5 and 4.4 for J024801-033258 and J045408-030014, respectively \citep{2014ApJ...792...76B,2016ApJ...817...11H}.
The absolute magnitudes, $M_{\rm UV}$, are corrected for the magnification factors, while the observed magnitudes are not.}
\vspace{-6pt}\tablenotetext{\small }{[references] 1) \citet{2022PASJ...74..247A}, 2) Ono et al. in prep, 3) \citet{2016ApJ...828...26M}, 4) \citet{2020ApJ...896...93H}, 5) \citet{2018PASJ...70S..35M}, 6) \citet{2012MNRAS.422.1425C}, 7) \citet{2018ApJS..235...36M}, 8) Sawicki et al. in prep, 9) \citet{2013MNRAS.428.1281J}, 10) \citet{2015ApJ...807..180W}, 11) \citet{2018MNRAS.479.1180M}, 12) \citet{2014ApJ...792...76B}, 13) \citet{2016ApJ...817...11H}, 14) \citet{2018ApJS..237....5M}, 15) \citet{2012ApJ...760..128M}, 16) \citet{2022ApJS..258...11W}, 17) \citet{2019ApJ...883..183M}, 18) \citet{2018Natur.553..178S}, 19) \citet{2022ApJ...928...31S}, 20) \citet{2015Natur.519..327W}, 21) \citet{2022ApJ...929..161W}, 22) \citet{2022ApJ...934...64A}, 23) \citet{2014MNRAS.440.2810B}, 24) \citet{2019PASJ...71...71H}, 25) \citet{2022ApJ...931..160B}, 26) \citet{2023MNRAS.518.6142A}, 27) \citet{2015Natur.522..455C}, 28) \citet{2020ApJS..247...61F}}
}
\vspace{-22pt}
\end{deluxetable*}

%
%
%
%
%

We fit the broad-band photometry from the catalogs to estimate $\beta_{\rm UV}$ and $M_{\rm UV}$, assuming a simple power-law spectral shape described by $f_{\lambda}\propto\lambda^{\beta_{\rm UV}}$ with a truncation at $\lambda_{\rm rest}=1216$\AA\ corresponding to the intergalactic medium (IGM) absorption by {\sc Hi} gas.
We take into account the contribution of the Ly$\alpha$ emission line if the line is clearly detected in the spectroscopic observations.
The brightness of the Ly$\alpha$ line is obtained from the literature or calculated from spectra following the methodologies in \citet{2020ApJ...896...93H}.
We fit model UV SEDs created with the logarithmic ranges of [-0.5, 0.5] for the normalization flux in the $y$-band and [-5, 3] for the UV slope $\beta_{\rm UV}$.
From each set of parameters, we derive $M_{\rm UV}$ by interpolating the model spectrum to the rest frame of 1500\AA.
The best-fit value and associated 1$\sigma$ uncertainty of $\beta_{\rm UV}$ and $M_{\rm UV}$ are derived as the value showing the minimum $\chi^2$ and the deviation at which $\Delta\chi^2=1$ from that the minimum $\chi^2$ value, respectively.
We find a typical reduced $\chi^2$ value is $\chi^2_{\nu}=3.0$, except for the sources covered in two photometric bands.

%
%
%
%
%
%
\begin{figure}[t]
\begin{center}
\epsscale{1.15}
\includegraphics[width=9cm,bb=5 5 400 400, trim=0 1 0 0cm]{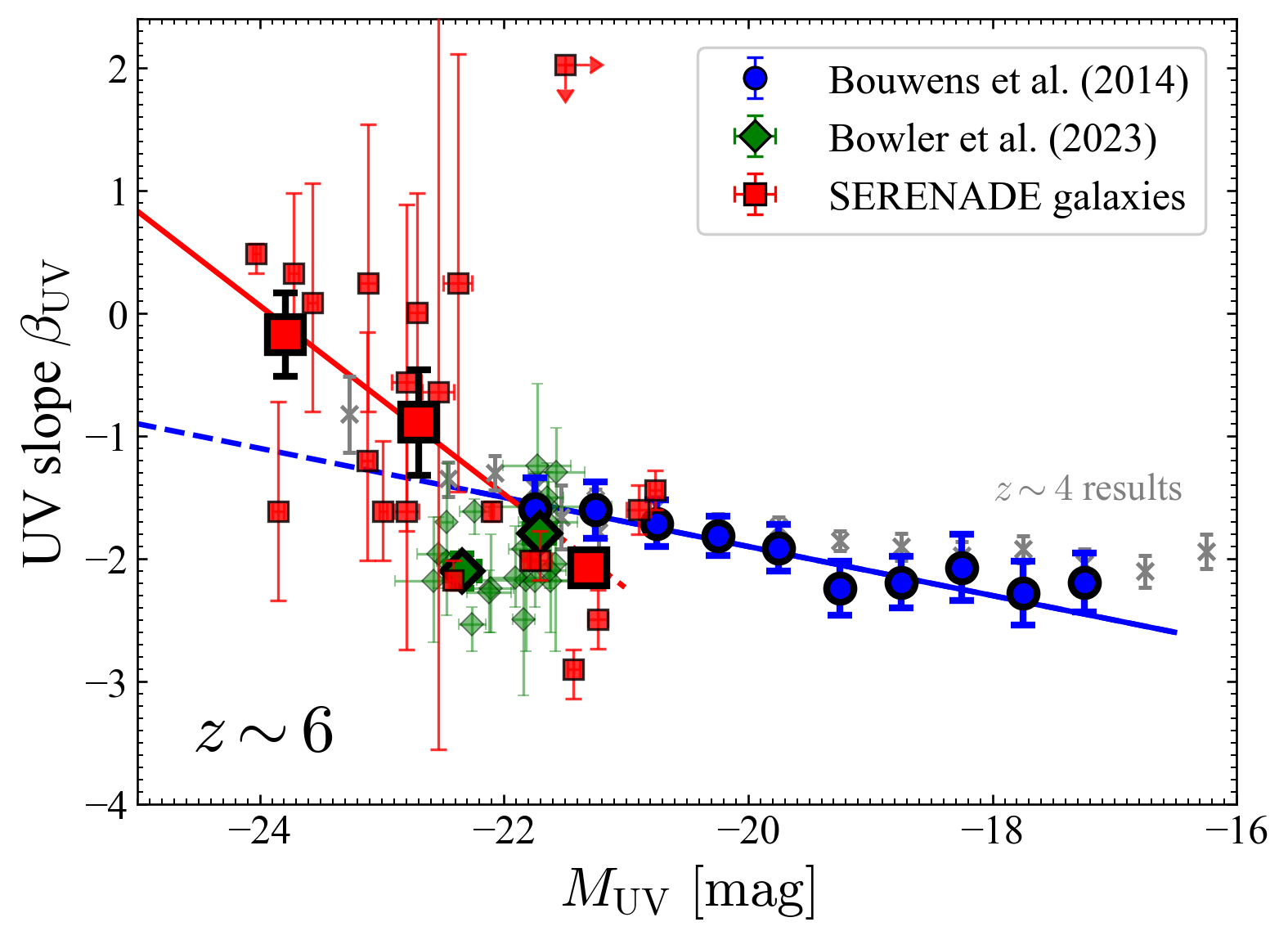}
\caption{Relationship between the rest-frame UV absolute magnitude and UV slope $\beta_{\rm UV}$. The SERENADE galaxies (red squares) are typically brighter than the binned results of {\it HST}-identified galaxies \citep[blue points;][]{2014ApJ...793..115B}. Our bright galaxies with $M_{\rm UV}<-22.0\,{\rm mag}$ generally exhibit redder slopes than the extrapolation of the relation constrained with faint galaxies in \citet{2014ApJ...793..115B}. Such a trend at the bright end has also been reported in previous studies at $z\sim4$ \citep[gray crosses;][see also, \citealt{2019PASJ...71...51Y}]{2011ApJ...733...99L,2014ApJ...793..115B}.}
\label{fig:Muv_beta}
\end{center}
\end{figure}

\subsection{$M_{\rm UV}$ vs $\beta_{\rm UV}$ relationship}\label{subsec:Muv_beta}
In Figure \ref{fig:Muv_beta}, we plot individual galaxy measurements of $M_{\rm UV}$-$\beta_{\rm UV}$ and representative average data points by binning our galaxies into three $M_{\rm UV}$ bins of [-24.5:-23.5], [-23.5:-22.0] and [-22.0:-20.5]. 
The error bar represents a $1\sigma$ standard deviation of the sample.
This result is consistent with the previous results, supporting the idea that more UV-luminous star-forming galaxies have redder UV-slope $\beta_{\rm UV}$ \citep[e.g.,][]{2010ApJ...712.1070R,2014ApJ...793..115B,2019PASJ...71...51Y}.
The UV slope could be redder than the extrapolated relationship from \citet{2014ApJ...793..115B} (hereafter B14) at $M_{\rm UV}<-22.0$.
Such a bending trend at the bright end has been reported in previous studies \citep[e.g.,][]{2011ApJ...733...99L}.
Our best-fit $M_{\rm UV}-\beta_{\rm UV}$ relationship normalized in  $M_{\rm UV}=-22$ is represented as $\beta_{\rm UV}=-(0.76\pm0.14)(M_{\rm UV}+22)-(1.51\pm0.15)$.
In contrast, \citet{2023arXiv230917386B} show potential flattening or turnover at $M_{\rm UV}<-21.5$, which may originate from a low dust obscuration due to clumpy geometry of the REBELS galaxies.

The Lyman-break selection technique could introduce biases favoring blue galaxies (i.e., small $\beta_{\rm UV}$).
The typical $i$-dropout selection criteria are composed of not only a red $i$-$z$ color to identify the Lyman-break but also a blue $z$-$y$ color coming from blue UV continuum \citep[e.g.,][see also \citealt{1999ApJ...519....1S}]{2012ApJ...754...83B,2014ApJ...793..115B,2022ApJS..259...20H}.
For example, the selection criteria used in \citet{2012ApJ...754...83B} select galaxies with $\beta_{\rm UV}\lesssim0.5$ at $z\sim5.8$--6.3. 
Given that our target galaxies are collected from a variety of surveys, potential biases could permeate our results.
While it is difficult to evaluate the biases on $M_{\rm UV}-\beta_{\rm UV}$ relationship, we test the potential bias by limiting galaxies selected from Subaru High-$z$ Exploration of Low-luminosity Quasars (SHELLQs) survey and follow-up spectroscopic observations.
As the galaxies from the SHELLQs survey are selected without any color criteria on the UV-slope $\beta_{\rm UV}$ \citep[see Figure 11 in][]{2018PASJ...70S..35M}, the galaxy sample offers a more homogeneous selection compared to the galaxies culled from surveys external to the SHELLQs.\footnote[1]{As the selection of $z\sim6$ quasars in the SHELLQs survey are composed of the red color originated from Lyman break ($i-z>1.5$) and the compactness, there may cause other bias apart from the UV slope \citep[see][for more detail]{2016ApJ...828...26M}.}
As a result, there is no major change if we calculate the average values only with the galaxies from the SHELLQs survey.
Consequently, we conclude that the overall trend is authentic although enlarging the sample size is imperative to draw robust conclusions.
Note that an average in the range $-22\leq M_{\rm UV}$ may be biased to bluer $\beta_{\rm UV}$ since all five galaxies in this range originate from inhomogeneous surveys.

\subsection{Dust continuum fluxes}\label{subsec:dustflux}
In this section, we describe the detection and measurement of the dust continuum fluxes.
First, we make data cubes with the taper scale of $1.0''$ and apply a single Gaussian fitting to the spectra extracted at the phase center to identify \cii\ 158\,$\mu$m or \oiii\ 88\,$\mu$m emission lines.
If any emission line feature is identified (i.e., Gaussian fitting converges), we mask a frequency range that is three times wider than the line velocity dispersion using {\sc CASA/mstransform}.
For cases with non-detection, we exclude the frequencies at $\pm0.8\,{\rm GHz}$ (equivalent to $\pm500\,{\rm km}\,{\rm s}^{-1}$) from the expected central frequency from their prior redshift to ensure eliminating potential emission line contamination to the dust continuum.
Subsequently, we generate continuum images with {\sc CASA/tclean}.
To recover all possible extended structures, we reconstruct images with $uv$ taper of $0.25''$, $0.5''$, $1.0''$, $2.0''$, and $3.0''$ in addition to the naturally-weighted image without tapering.
We obtain peak flux density within a $0.5''$ radius from each galaxy's central position, under the assumption that dust continuum emission is included within a single beam.
Noise levels are estimated in the images without primary beam corrections as the root-mean-square (RMS) of pixel values within the Field-of-View (FoV) where the primary beam correction value exceeds 0.5.
As the fiducial signal-to-noise ratio (S/N) of each galaxy, we adopt the highest S/N among the images.
Galaxies with fiducial ${\rm S/N}>3.5$ in any band are considered as continuum detection.

\begin{figure*}[htbp]
\begin{center}
\epsscale{1.15}
\includegraphics[width=16cm,bb=0 0 1000 650, trim=0 1 0 0cm]{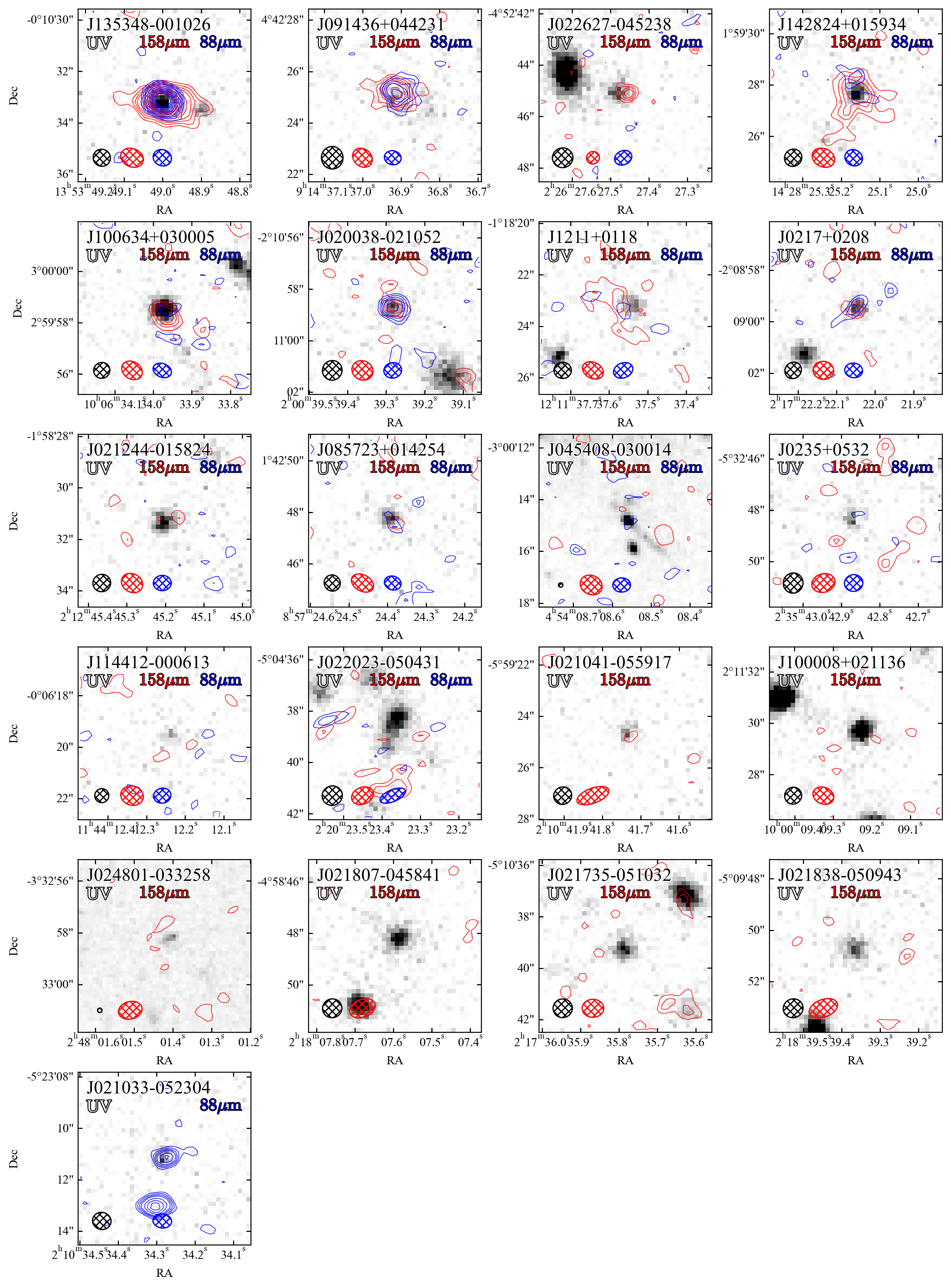}
\caption{Thumbnails of the SERENADE galaxies. The red and blue contours indicate rest-frame $158\mu{\rm m}$ and $88\mu{\rm m}$ dust continua, respectively. The contours are shown every 1$\sigma$ from 3$\sigma$ to 5$\sigma$ and every 2$\sigma$ from 5$\sigma$. The background images are {\it HST} $Y_{105}J_{125}H_{160}$-combined images for J024801-033258 amd J045408-030014 and Subaru/HSC $z$-band images for others. The source ID and beam sizes in each wavelength are also shown.}
\label{fig:thumnails}
\end{center}
\end{figure*}

Figure \ref{fig:thumnails} shows the thumbnails of the SERENADE galaxies. 
The contours of the natural-weighted dust continuum in the rest-frame $158\mu{\rm m}$ (red) and $88\mu{\rm m}$ (blue) are overlayed on the HSC $z$-band or {\it HST} $Y_{105}J_{125}H_{160}$ combined images.
Of the 14 galaxies observed in both Band-6 and Band-8, six galaxies are detected in both the rest-frame $158\mu{\rm m}$ and $88\mu{\rm m}$, two galaxies are only detected in the rest-frame $158\mu{\rm m}$ dust continuum, and six galaxies are detected neither $158\mu{\rm m}$ nor $88\mu{\rm m}$.
There are also six galaxies observed solely in Band-6 and one in Band-8.

We measure the total flux density at multiple frequencies with consideration for the different spatial resolution between the rest-frame $158\mu{\rm m}$ and $88\mu{\rm m}$ continuum.
To ensure that the total fluxes are measured and to achieve uniform measurements across frequencies, we employ the following three methods to calculate fiducial flux values; (1) For the galaxies detected in both rest-frame $158\mu{\rm m}$ and $88\mu{\rm m}$, we execute {\sc CASA/imfit}, (2) For those detected only in $158\mu{\rm m}$, we get the peak flux density and RMS in images with $uv$ taper of $1.0''$ following the S/N calculations described above, (3) For the galaxies not detected in any bands, we calculate upper limits from the naturally weighted images.
The reason behind method (2) is to ensure uniformity in the calculation of the RMS levels between the detected and undetected frequencies.
To avoid the flux misestimation in {\sc imfit} due to the mismatch of dirty and clean beams, we apply modest tapers for some sources before running {\sc imfit}.
From the comparison between the fiducial flux value calculated above and the total flux derived from the visibility analysis presented in the Appendix, we find that calculated fiducial fluxes, in any case, are consistent with those from the visibility analysis.
We also find that {\sc CASA/imfit} achieves better flux recovering than that from the images with the multiple taper scales. 
These comparisons are shown in Figure \ref{fig:fluxcomp} in Appendix \ref{appendix:fluxcomp}.



%
%
%
%
%
%

\subsection{Stacking analysis}\label{subsec:stacking}
\begin{figure}[t]
\begin{center}
\epsscale{1.15}
\includegraphics[width=7.5cm,bb=5 5 400 800, trim=0 1 0 0cm]{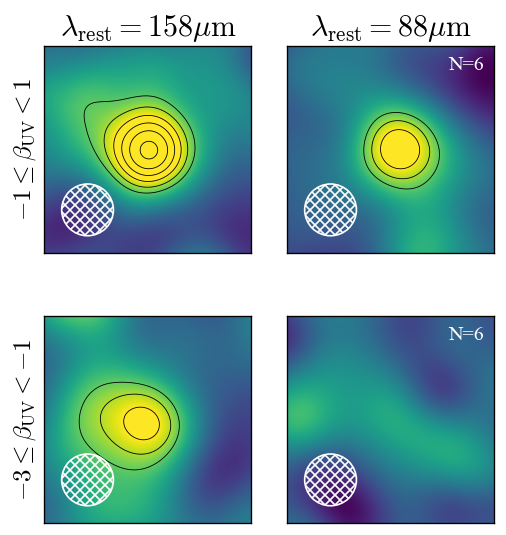}
\caption{$6''\times6''$ cutout of the stacked images in both rest-frame $158\mu{\rm m}$ and $88\mu{\rm m}$. We show the number of galaxies included in each bin and the representative beam sizes of each stacked image. The contour levels are every 1$\sigma$ started from 3$\sigma$.}
\label{fig:simage}
\end{center}
\end{figure}

In this section, we stack the dust continuum of the SERENADE galaxies at $z\sim6$ to obtain an average representation that is not biased by individual detection.
We perform the stacking of both rest-frame 88$\mu{\rm m}$ and 158$\mu{\rm m}$ dust continuums from galaxies in two distinct $\beta_{\rm UV}$ bins of $\beta_{\rm UV}=[-3,-1]$ and $[-1,1]$, where each bin include approximately an equal number of galaxies.
Only galaxies observed in both Band-6 and Band-8 are used in the stacking analysis for the sake of homogeneity.
Additionally, we exclude two galaxies (J020038-021052 and J024801-033258) because J020038-021052 do not have $M_{\rm UV}$ estimation and J024801-033258 is lensed.
The resulting stacked images are shown in Figure \ref{fig:simage}.
Given the variable nature of the ALMA beam sizes, we construct continuum images with a round $2\farcs0$ beam by applying {\sc CASA/imsmooth} for the images with a prior $1''$ taper.
For the reference position of the stacking, we use central positions derived in the rest-frame UV continuum (i.e., phase center of the ALMA observations).
We note that spatial offsets identified in individual galaxies between UV and IR do not significantly impact the flux measurement since the spatial offsets are less than $0.35''$.
Then we also compute the weighted average with the following weight based on the UV luminosity of $i$-th galaxy in the sample:

\begin{equation}
w_i = \frac{L_{\rm UV}^{\rm mean}}{L_{{\rm UV},i}}
\end{equation}

\noindent This reciprocal weighting with respect to $L_{\rm UV}$ allows us to avoid bias towards bright galaxies.
Finally, we measure fluxes with the same procedure described in Section \ref{subsec:dustflux}.
We find that there is no major difference when we use the weighted median instead of the weighted average.

%
%
%
%
%
%

\subsection{$T_{\rm dust}$ and $L_{\rm IR}$ estimation}\label{subsec:mcmc}
We fit a modified blackbody (MBB) to constrain the properties of the dust emission.
The MBB profile is mainly determined by three parameters, dust temperature ($T_{\rm dust}$), dust mass ($M_{\rm dust}$), and emissivity of the dust grain ($\beta_{\rm dust}$).
The general formulation of the observed MBB flux density at $\nu_{\rm obs}$ is as follows:

\begin{equation}\label{eq2}
F_{\nu_{\rm obs}}=\left(\frac{1+z}{d_{\rm L}^2}\right)\left(\frac{1-e^{-\tau_{\nu}}}{\tau_{\nu}}\right)M_d\kappa_{\nu}B_{\nu}(T_{\rm dust}).
\end{equation}

\noindent Here $d_{\rm L}$ and $\tau_{\nu}$ are the luminosity distance at redshift $z$ and the optical depth of the dust grain, respectively.
$B_{\nu}(T_{\rm dust})$ represents the blackbody radiation at the temperature of $T_{\rm dust}$.
The absorption coefficient $\kappa_{\nu}$ is assumed by $\kappa_{\nu}=\kappa_{\ast}(\nu/\nu_{\ast})^{\beta_{\rm dust}}$ with the nomalization of Milky Way value, [$\kappa_{\ast}$, $\nu_{\ast}$]=[10.41 cm$^{-2}$g$^{-1}$, 1900 GHz] \citep[e.g.,][]{2021MNRAS.508L..58B,2023MNRAS.518.6142A,2022MNRAS.512...58F}.
The dust optical depth is defined by the net absorption over the line of sight and represented as $\tau_{\nu}=\kappa_{\nu}\Sigma_{\rm dust}$, where $\Sigma_{\rm dust}$ represents the surface dust mass density.
$\tau_{\nu}$ is also commonly written as $(\nu/\nu_0)^{\beta}=(\lambda_0/\lambda)^{\beta}$, where $\nu_0$ is a frequency at which the optical depth may exceed unity (e.g., around $\lambda_0=100$--200$\mu{\rm m}$ for dusty star-forming galaxies; \citealt{2012MNRAS.425.3094C}). 

It is important to take into account the cosmic microwave background (CMB) effect, especially in the high-$z$ universe since the CMB temperature increases with redshift as $T_{{\rm CMB},z}=T_{{\rm CMB},z=0}\times(1+z)$.
The CMB affects the $T_{\rm dust}$ measurements in two ways;
(1) additional heating source of dust grains within galaxies (2) background radiation against which we observe the submillimeter fluxes from galaxies \citep{2013ApJ...766...13D}.
After considering these effects following \citet{2013ApJ...766...13D}, the equation (\ref{eq2}) becomes

\begin{equation}
F_{\nu_{\rm obs}}=\left(\frac{1+z}{d_{\rm L}^2}\right)\left(\frac{1-e^{-\tau_{\nu}}}{\tau_{\nu}}\right)M_d\kappa_{\nu}[B_{\nu}(T_{\rm dust})-B_{\nu}(T_{\rm CMB})].
\end{equation}

\noindent In the optically thin limit ($\nu_0\rightarrow\infty$, $\lambda_0\rightarrow0$ and thus $\tau_{\nu}\ll1$), the term representing the effect of the optical depth ($1-e^{-\tau_{\nu}}/\tau_{\nu}$) is asymptotically unity.
Therefore the formula of the optically thin MBB becomes 

\begin{equation}\label{eq4}
F_{\nu_{\rm obs}}=\left(\frac{1+z}{d_{\rm L}^2}\right)M_d\kappa_{\nu}[B_{\nu}(T_{\rm dust})-B_{\nu}(T_{\rm CMB})].
\end{equation}


\noindent In this work, we use equation (\ref{eq4}) to estimate $T_{\rm dust}$, $\beta_{\rm dust}$, $M_{\rm dust}$ in the same manner as previous studies.
We note that the resulting $L_{\rm IR}$ does not change under the assumption of the optically thick MBB with $\lambda_0=100\mu{\rm m}$, whereas $T_{\rm dust}$ becomes higher than those with the optically thin assumption.

\tabcolsep = 0.1cm
%
%
%
%
%
%
\begin{deluxetable*}{cccccccccc}
\tablecaption{Summary of the parameters related to rest-frame FIR emission \label{tab:tab2}}
\tablewidth{0pt}
\tablehead{\colhead{ID} & 
\colhead{$S_{158\mu{\rm m}}$} & 
\colhead{$S_{88\mu{\rm m}}$} &
\colhead{$S_{205\mu{\rm m}}$} & 
\colhead{$S_{122\mu{\rm m}}$} & 
\colhead{$S_{110\mu{\rm m}}$} & 
\colhead{$S_{52\mu{\rm m}}$} &
\colhead{$T_{\rm dust}$\textsuperscript{\mbox{*}}} & 
\colhead{$\log L_{\rm IR}$} & 
\colhead{$\log M_{\rm dust}$} \\
 & ${\rm [mJy]}$ & ${\rm [mJy]}$ & ${\rm [mJy]}$ & ${\rm [mJy]}$ & ${\rm [mJy]}$ & ${\rm [mJy]}$ & ${\rm [K]}$ & $[L_{\odot}]$ & $[M_{\odot}]$ 
}
\startdata
\multicolumn{10}{c}{{\bf SERENADE}} \\
J020038-021052 & $0.24\pm0.08$ & $0.75\pm0.10$ & - & - & - & - & $56.4_{-24.1}^{+28.7}$ & $11.9_{-0.2}^{+0.7}$ & $6.8_{-0.3}^{+0.6}$ \\
J021033-052304 & - & $1.93\pm0.25$ & - & - & - & - & ({\it 45}) & $12.3_{-0.1}^{+0.1}$ & $9.0_{-0.1}^{+0.1}$ \\
J021041-055917 & $\leq0.14$ & - & - & - & - & - & ({\it 45}) & $\leq11.6$ & $\leq7.3$ \\
J021244-015824 & $\leq0.17$ & $\leq0.62$ & - & - & - & - & ({\it 45}) & $\leq11.7$ - & $\leq7.4$ \\
J021735-051032 & $\leq0.07$ & - & - & - & - & - & ({\it 45}) & $\leq11.3$ & $\leq7.0$ \\
J021807-045841 & $\leq0.06$ & - & - & - & - & - & ({\it 45}) & $\leq11.2$ & $\leq6.9$ \\
J021838-050943 & $\leq0.06$ & - & - & - & - & - & ({\it 45}) & $\leq11.2$ & $\leq7.0$ \\
J022023-050431 & $\leq0.16$ & $\leq0.83$ & - & - & - & - & ({\it 45}) & $\leq11.7$ - & $\leq7.4$ \\
J022627-045238 & $0.18\pm0.03$ & $\leq0.50$ & - & - & - & - & $27.4_{-7.6}^{+20.3}$ & $11.2_{-0.2}^{+0.5}$ & $7.6_{-0.6}^{+0.7}$ \\
J024801-033258 & $\leq0.03$ & - & - & - & - & - & ({\it 45}) & $\leq11.0$ & $\leq6.7$ \\
J045408-030014 & $\leq0.11$ & $\leq0.26$ & - & - & - & - & ({\it 45}) & $\leq11.5$ - & $\leq7.2$ \\
J085723+014254 & $\leq0.13$ & $\leq0.33$ & - & - & - & - & ({\it 45}) & $\leq11.5$ - & $\leq7.2$ \\
J091436+044231 & $0.88\pm0.11$ & $4.35\pm0.59$ & - & - & - & - & $75.6_{-23.8}^{+24.4}$ & $13.4_{-0.6}^{+0.2}$ & $7.4_{-0.1}^{+0.3}$ \\
J100008+021136 & $\leq0.04$ & - & - & - & - & - & ({\it 45}) & $\leq11.1$ & $\leq6.8$ \\
J100634+030005 & $0.38\pm0.07$ & $0.69\pm0.26$ & - & - & - & - & $30.1_{-7.8}^{+16.9}$ & $11.6_{-0.2}^{+0.3}$ & $7.9_{-0.5}^{+0.6}$ \\
J114412-000613 & $\leq0.13$ & $\leq0.46$ & - & - & - & - & ({\it 45}) & $\leq11.6$ & $\leq7.3$ \\
J135348-001026 & $2.74\pm0.16$ & $7.97\pm0.23$ & - & - & - & - & $44.1_{-10.5}^{+16.9}$ & $12.9_{-0.1}^{+0.2}$ & $8.6_{-0.4}^{+0.3}$ \\
J142824+015934 & $1.10\pm0.18$ & $4.10\pm1.30$ & - & - & - & - & $49.3_{-14.8}^{+37.0}$ & $12.9_{-0.4}^{+0.6}$ & $7.6_{-0.2}^{+0.5}$ \\
\hline
\multicolumn{10}{c}{{\bf SERENADE pilot} \citep{2020ApJ...896...93H}} \\
J1211+0118 & $0.14\pm0.03$ & $\leq0.62$ & - & $0.23\pm0.03$ & - & - & $37.3_{-13.5}^{+36.1}$ & $11.3_{-0.2}^{+0.9}$ & $6.7_{-0.2}^{+0.8}$ \\
J0217+0208 & $0.14\pm0.03$ & $0.53\pm0.14$ & - & $0.22\pm0.03$ & - & - & $56.8_{-18.9}^{+36.0}$ & $12.4_{-0.7}^{+0.4}$ & $6.6_{-0.1}^{+0.5}$ \\
J0235+0532 & $\leq0.07$ & $\leq0.28$ & - & $0.09\pm0.02$ & - & - &$53.2_{-21.0}^{+43.7}$ & $11.9_{-0.3}^{+1.1}$ & $6.1_{-0.3}^{+0.6}$ \\
\hline
\multicolumn{10}{c}{$z\sim7$ {\bf LBGs} \citep{2019PASJ...71...71H,2021MNRAS.508L..58B,2022ApJ...934...64A,2022MNRAS.515.1751W,2023MNRAS.518.6142A}} \\
COS-2987030247 & $\leq0.02$ & $\leq0.19$ & - & - & - & - & ({\it 45}) & $\leq10.9$ & $\leq6.6$ \\
COS-3018555981 & $0.05\pm0.01$ & $\leq0.20$ & - & - & - & - & $30.1_{-7.3}^{+23.7}$ & $10.9_{-0.2}^{+0.5}$ & $7.0_{-0.5}^{+0.7}$ \\
UVISTA-Z-001 & $0.07\pm0.01$ & $0.23\pm0.07$ & - & - & - & - & $31.4_{-9.1}^{+35.2}$ & $11.3_{-0.3}^{+0.7}$ & $6.7_{-0.5}^{+0.9}$ \\
UVISTA-Z-007 & $\leq0.05$ & $\leq0.30$ & - & - & - & - & ({\it 45}) & $\leq11.3$ & $\leq7.0$ \\
UVISTA-Z-019 & $0.07\pm0.01$ & $\leq0.83$ & - & - & - & - & $32.5_{-9.9}^{+52.8}$ & $12.2_{-1.4}^{+0.2}$ & $6.5_{-0.2}^{+0.6}$ \\
A1689-zD1 & $1.09\pm0.17$ & $2.68\pm0.19$ & - & - & $2.04\pm0.14$ & $1.84\pm0.43$ & $45.3_{-6.6}^{+8.5}$ & $11.4_{-0.1}^{+0.1}$ & $7.1_{-0.2}^{+0.2}$ \\
B14-65666 & $0.13\pm0.03$ & $0.47\pm0.14$ & - & - & $0.22\pm0.01$ & - & $51.3_{-25.1}^{+33.8}$ & $11.8_{-0.5}^{+0.7}$ & $6.7_{-0.2}^{+0.6}$ \\
REBELS-12 & $0.05\pm0.01$ & $\leq0.32$ & - & - & - & - & $34.9_{-11.3}^{+47.8}$ & $11.1_{-0.2}^{+1.0}$ & $6.3_{-0.3}^{+0.7}$ \\
REBELS-25 & $0.22\pm0.01$ & $0.69\pm0.03$ & - & - & - & - & $41.2_{-9.0}^{+20.2}$ & $11.9_{-0.1}^{+0.2}$ & $7.5_{-0.3}^{+0.4}$ \\
REBELS-38 & $0.18\pm0.02$ & $0.51\pm0.10$ & - & - & - & - & $43.6_{-14.2}^{+19.4}$ & $11.7_{-0.2}^{+0.4}$ & $7.4_{-0.5}^{+0.3}$ \\
\hline
\multicolumn{10}{c}{$z\sim5.5$ {\bf LBGs} \citep{2020ApJS..247...61F}} \\
HZ4 & $0.27\pm0.03$ & - & $0.11\pm0.01$ & - & $0.50\pm0.07$ & - & $47.9_{-14.3}^{+35.4}$ & $12.0_{-0.3}^{+0.6}$ & $6.9_{-0.3}^{+0.4}$ \\
HZ6 & $0.42\pm0.04$ & - & $0.27\pm0.04$ & - & $0.58\pm0.09$ & - & $25.7_{-7.8}^{+7.6}$ & $11.4_{-0.1}^{+0.2}$ & $8.3_{-0.5}^{+0.5}$ \\
HZ9 & $0.59\pm0.09$ & - & $0.31\pm0.02$ & - & $1.18\pm0.09$ & - & $42.5_{-11.5}^{+29.4}$ & $12.1_{-0.2}^{+0.5}$ & $7.5_{-0.3}^{+0.4}$ \\
HZ10 & $1.87\pm0.08$ & - & $0.69\pm0.03$ & - & $3.39\pm0.18$ & - & $33.4_{-5.4}^{+10.0}$ & $12.5_{-0.1}^{+0.2}$ & $8.6_{-0.4}^{+0.2}$ \\
\hline
\multicolumn{10}{c}{{\bf stack}} \\
$-3<\beta_{\rm UV}<-1$ & $0.21\pm0.04$ & $\leq0.63$ & - & - & - & - & $50.4_{-9.5}^{+24.6}$ & $12.2_{-0.2}^{+0.7}$ & $7.4_{-0.2}^{+0.3}$ \\
$-1\leq\beta_{\rm UV}<1$ & $0.36\pm0.01$ & $1.24\pm0.03$ & - & - & - & - & $26.0_{-6.5}^{+14.5}$ & $11.3_{-0.2}^{+0.4}$ & $7.8_{-0.6}^{+0.7}$ \\
\enddata

\tablecomments{
\vspace{-6pt}\tablenotetext{\small \ast}{For the galaxies whose dust temperatures are not constrained due to insufficient observations or non-detections, we assume the representative dust temperature of the SERENADE galaxies ($T_{\rm dust}=45\,{\rm K}$).}
\vspace{-22pt}
}
\end{deluxetable*}
%
%
%
%
%

%
%
%
%
%
%
\begin{figure*}[htbp]
\begin{center}
\epsscale{1.15}
\includegraphics[width=16.5cm,bb=0 0 1000 650, trim=0 1 0 0cm]{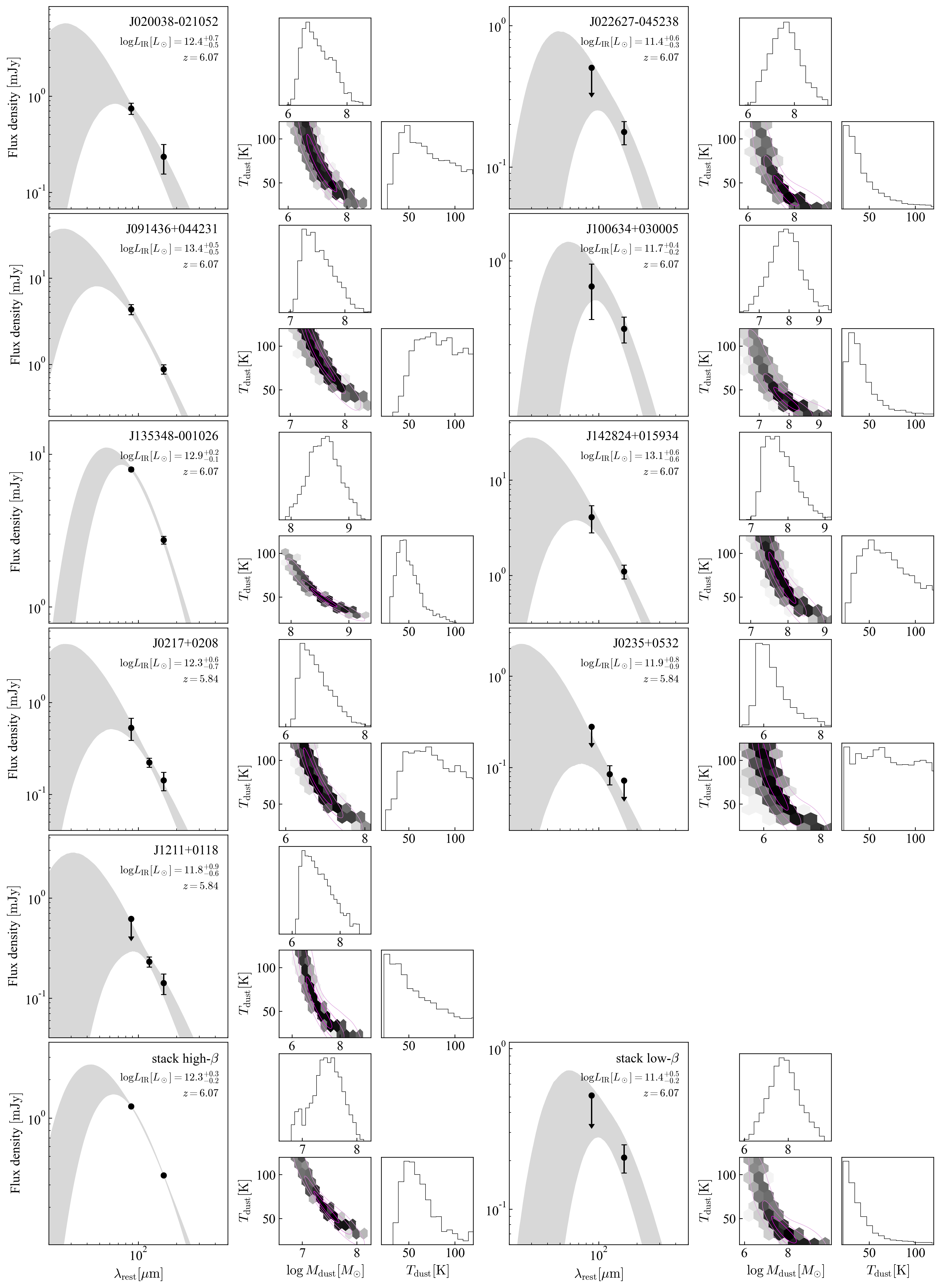}
\caption{Optically thin MBB fitting results of the SERENADE galaxies. The left panels show the 1$\sigma$ confidence interval of the MBB profiles as a function of the rest-frame wavelength (shaded area). The observed fluxes and 3$\sigma$ upper limits are shown in the black circles. The right panels illustrate the posterior distributions of the MCMC procedure with 1-,2- and 3-$\sigma$ contours. The best-fit and 1$\sigma$ uncertainties of $L_{\rm IR}$ are shown inside the left panels.}
\label{fig:Tdfitting}
\end{center}
\end{figure*}

We use the Markov Chain Monte Carlo (MCMC) algorithm to fit the MBB profiles by utilizing the {\sc emcee} library. 
The fitting results are shown in Figure \ref{fig:Tdfitting}.
We adopt a logarithmically uniform prior on the dust masses and dust temperatures with a range of $\log_{10}(M_{\rm dust}/M_{\odot})\in[4,10]$ and $T_{\rm dust}/{\rm K}\in[T_{{\rm CMB},z},120]$, respectively.
Note that there is no significant impact on our conclusion if we change the upper edge of the prior distribution of $T_{\rm dust}$ to $150\,{\rm K}$.
Since the sampling range of the FIR SED is not enough to constrain $\beta_{\rm dust}$, we adopt a Gaussian prior with a mean value of $\langle\beta_{\rm dust}\rangle=1.8$ and a standard deviation of $\sigma_{\beta_{\rm dust}}=0.5$ to take the $\beta_{\rm dust}$ uncertainties into account to those of $T_{\rm dust}$, $M_{\rm dust}$ and the resulting $L_{\rm IR}$.
This value is consistent with that of Milky Way \citep{2014A&A...571A..11P}, the local galaxies \citep{2014MNRAS.440..942C,2019MNRAS.489.4389L}, local ULIRG \citep{2018MNRAS.475.2097C}, and high-z galaxies \citep{2021ApJ...919...30D,2023MNRAS.523.3119W}.
The results do not strongly depend on the prior distribution of $\beta_{\rm dust}$ as there is no significant difference if we apply a uniform prior of $\beta_{\rm dust}\in[1.5,2.5]$\footnote{In the case of the uniform prior, the fitting results tend to prefer slightly lower $\beta_{\rm dust}$ and $T_{\rm dust}$ values compared with those under the gaussian prior}.
We fit the optically thin MBB model to the measured flux densities.
In the case that the measured ALMA fluxes in any bands are upper limit, we adopt the methodology described in \citet{2012PASP..124.1208S}, which accounts for the Gaussian noise distribution of the probability distribution.
In a nutshell, the probability of a model flux $S_{\rm model}$ can be calculated by integrating a Gaussian distribution, which has a center of $S_{\rm model}$ and dispersion of $\sigma_{S_{\rm model}}$, from negative infinity to the 3-$\sigma$ value of the upper limits.
For instance, the probabilities taking model fluxes of $S_{\rm model}=S_{1\sigma}$ or $S_{3\sigma}$ are penalized by $\sim$0.1\% or $\sim$50\% respectively.
Therefore the model fluxes above 3-$\sigma$ are not forbidden but the model fluxes much smaller than 3-$\sigma$ are preferred.
This is a reason why the resulting MBB sometimes yields model fluxes above the 3$\sigma$ upper limit (see Figure \ref{fig:Tdfitting}).
We compute best-fit values and 1$\sigma$ uncertainties from the modes with the highest posterior density intervals.

Figure \ref{fig:Tdfitting} shows the results of the MBB fittings for the SERENADE galaxies.
Our MBB fitting successfully constrains $T_{\rm dust}$, $M_{\rm dust}$ and $L_{\rm IR}$, while some galaxies such as J020038-021052 or J0235+0532 need observations at shorter wavelengths or deeper observations are necessary to obtain more robust constraints on the parameters.
The fitting results are summarized in Table \ref{tab:tab2}.

\begin{figure}[t]
\begin{center}
\epsscale{1.15}
\includegraphics[width=8.5cm,bb=0 0 200 150, trim=0 1 0 0cm]{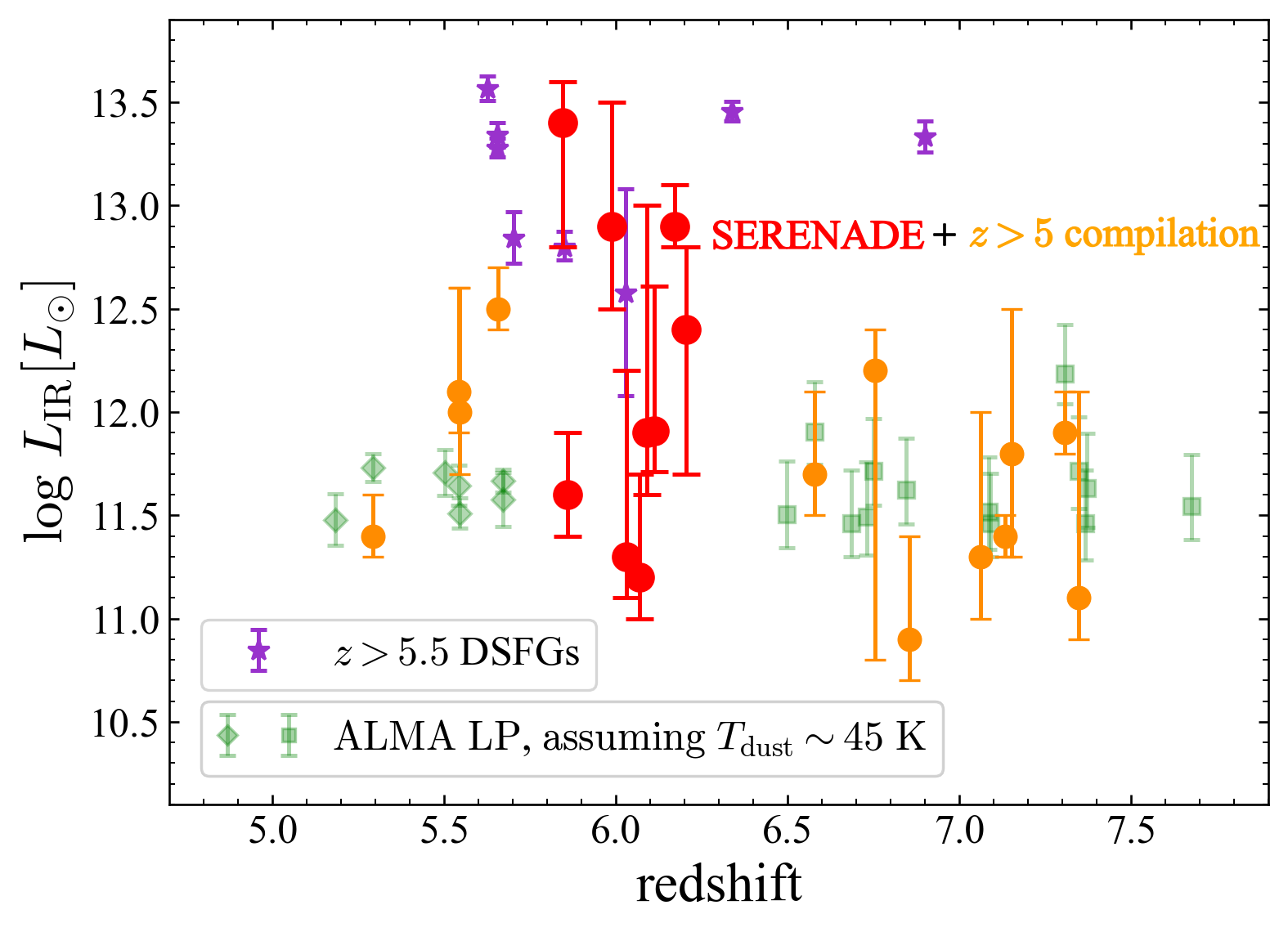}
\caption{IR luminosities of the SERENADE galaxies (red) and those at $5<z<8$ in the literature (orange) as a function of the redshift. We compare the results with galaxies reported in the ALMA large programs in which fixed dust temperatures are assumed \citep[green symbols;][]{2020ApJS..247...61F,2022MNRAS.515.3126I} and DSFGs at $z>5.5$ \citep[purple stars;][]{2013Natur.496..329R,2018NatAs...2...56Z,2018Natur.553...51M,2019ApJ...887...55C,2020ApJ...902...78R}. The definition of $L_{\rm IR}$ is uniform except for the HFLS3 at $z=6.34$ \citep[$L_{\rm FIR}$, 42.5-122.5\,$\mu$m;][]{2013Natur.496..329R}.}
\label{fig:zLIR}
\end{center}
\end{figure}

We compare the derived infrared luminosities ($L_{\rm IR}$,8--$1000\,\mu{\rm m}$) of the SERENADE galaxies with the literature in Figure \ref{fig:zLIR}.
For a fair comparison with the galaxies previously reported, we conduct the same flux measurements and MBB fittings for the galaxies at $5<z<8$ in the literature.
While our target galaxies were originally selected by their bright UV emission, the $L_{\rm IR}$ of the SERENADE galaxies spans a wide range, specifically from $\log L_{\rm IR}[L_{\odot}]\sim11.5$ comparable to normal galaxies at $z\sim7$ \citep[e.g.,][]{2022MNRAS.512...58F} to $\log L_{\rm IR}[L_{\odot}]\sim13.0$ similar to dusty starburst galaxies \citep[e.g.,][]{2013Natur.496..329R}.

%
%
%
%
%
%
%
%
%
%
%
%

%
%
%
%
%
%
\begin{figure}[t]
\begin{center}
\epsscale{1.15}
\includegraphics[width=8.5cm,bb=0 0 200 150, trim=0 1 0 0cm]{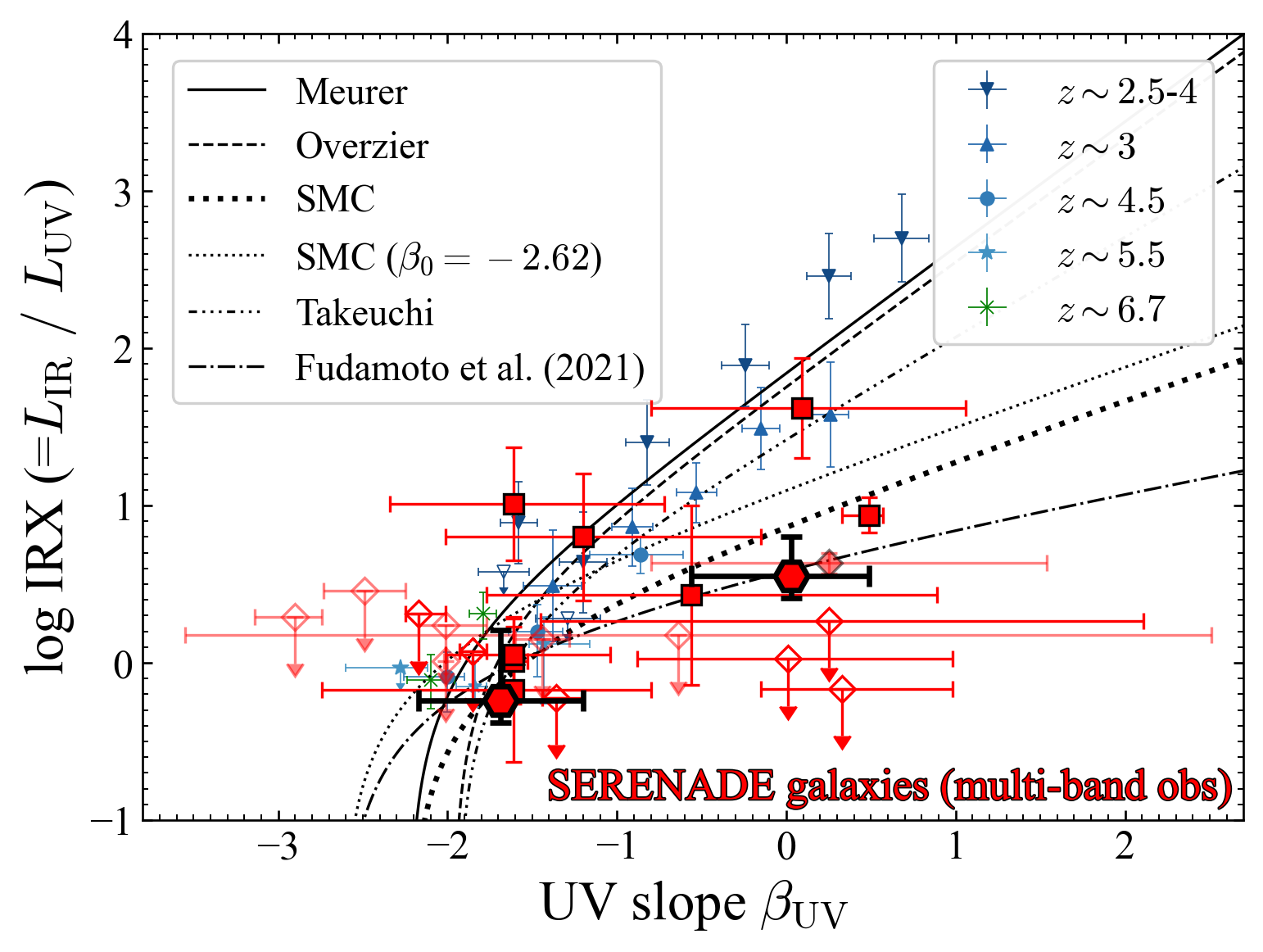}
\caption{IRX-$\beta_{\rm UV}$ relation. 
The filled red squares are the individual galaxies whose dust temperatures are constrained with multi-band observations. 
The filled pentagons show the results of the stacking analysis. 
The filled and open diamonds are galaxies whose dust temperatures are not constrained due to insufficient observations or non-detections. 
For these cases, we assume the dust temperature of $45\,{\rm K}$. 
For the stacking analysis, we use the galaxies observed in both ALMA bands. 
The blue and green markers represent the previous measurements of the galaxies at $z=2.5$--4 \citep{2020MNRAS.491.4724F}, $z\sim3$ \citep{2019A&A...630A.153A}, $z\sim4.5$ and $z\sim5.5$ \citep{2020A&A...643A...4F}, and $z\sim6.7$ \citep{2023arXiv230917386B}.
The best-fit relations from local \citep{1984A&A...132..389P,1999ApJ...521...64M,2000ApJ...533..682C,2003ApJ...594..279G,2011ApJ...726L...7O,2012ApJ...755..144T} and high-$z$ galaxies \citep[e.g.,][]{2016A&A...587A.122A,2018ApJ...853...56R,2020A&A...643A...4F} are also plotted. 
}
\label{fig:irx_beta}
\end{center}
\end{figure}
%
%
%
%
%
%

%
%
%
%
%
%
\section{Results}\label{sec:results}

\subsection{$\beta_{\rm UV}$ vs IRX relationship}\label{subsec:IRXbeta}
In Figure \ref{fig:irx_beta}, we plot the IRX-$\beta_{\rm UV}$ diagram of the SERENADE galaxies.
The results for the individually detected galaxies include uncertainties of the FIR SED.
Meanwhile, the upper limits of the undetected galaxies are calculated by assuming $T_{\rm dust}=45{\rm K}$ (a representative $T_{\rm dust}$ of SERENADE galaxies).
We find that our result from the stacking analysis supports a shallower IRX-$\beta_{\rm UV}$ relation than the Calzetti \citep{2000ApJ...533..682C} and Meurer \citep{1999ApJ...521...64M} relations.
Furthermore, our results prefer a shallower relation than the SMC extinction \citep[e.g.,][]{1998ApJ...508..539P,2003ApJ...594..279G,2018ApJ...853...56R} regardless of the intrinsic UV slope ($\beta_{\rm UV, int}=-2.23$ or $-2.62$) and match well the relation reported in \citet{2020A&A...643A...4F} with $\beta_{\rm UV, int}=-2.62$.
This shallower relation aligns with the findings of \citet{2018ApJ...853...56R}.
They expect that galaxies with lower stellar metallicities at high redshift ($Z_{\ast}\sim0.14Z_{\odot}$) favor a shallow IRX-$\beta_{\rm UV}$ relation with bluer intrinsic UV-slope $\beta_{\rm UV, int}=-2.62$.
Since our target luminous LBGs at $z\sim6$ are expected to have sub-solar stellar metallicities, such as $\sim0.4Z_{\odot}$ as presented in \citet{2020ApJ...902..117H}, our results have a good agreement with the prediction in \citet{2018ApJ...853...56R}.
As discussed in \citet{2023arXiv230917386B}, the difference from the results at $z\sim5$--7 \citep{2020A&A...643A...4F,2023arXiv230917386B} at $\beta_{\rm UV}=[-3:-1]$ mainly comes from the difference in $T_{\rm dust}$.
Our estimate of $T_{\rm dust}\sim25\,{\rm K}$ for the stacking sample with $\beta_{\rm UV}=[-3:-1]$ from multi-band observations is lower by $\sim20\,{\rm K}$ than the assumptions in these studies, and it leads to $\sim0.7\,{\rm dex}$ offset in the IRX.

We test potential biases in sample selection with the same procedure in Section \ref{subsec:stacking} and find that the result remains consistent even when selecting galaxies only from the SHELLQs survey.
Some studies have implied that the methodology for estimating $\beta_{\rm UV}$ could systematically change the IRX-$\beta_{\rm UV}$ relation. 
For instance, \citet{2019A&A...630A.153A} suggested that the $\beta_{\rm UV}$ estimated by power-law fitting ($\beta_{\rm UV, power}$) may be biased to redder values by $\Delta\beta_{\rm UV}\sim0.35$ than that derived by SED fitting ($\beta_{\rm UV, SED}$).
Furthermore, \citet{2015ApJ...806..259R} demonstrated that $\beta_{\rm UV}$ calculated within the rest-frame wavelength range of $\lambda_{\rm rest}=$1260\AA--1750\AA\ ($\beta_{\rm UV, narrow}$) might be redder than that within the range of $\lambda_{\rm rest}=$1260\AA--2600\AA\ ($\beta_{\rm UV, wide}$) especially in the spectrum with large $\beta_{\rm UV}$.
Given the analysis in \citet{2015ApJ...806..259R}, the redder $\beta_{\rm UV}$ bin of our galaxies, spanning $-0.5\lesssim\beta_{\rm UV}\lesssim0.5$, may intrinsically be bluer than the current estimation such as $\beta_{\rm UV}\sim-0.8$ since four of five galaxies within $-0.5\lesssim\beta_{\rm UV}\lesssim0.5$ have limited wavelength coverages of $\lambda_{\rm rest}=$1260\AA--1700\AA.
However, even after changing $\beta_{\rm UV}\sim0.0$ to $-0.8$, our results still favor a shallow IRX-$\beta_{\rm UV}$ relation such as $z\sim5.5$ results from \citet{2020A&A...643A...4F}.
To secure robust measurements of UV slopes, NIR observations that cover the rest-frame optical wavelengths at $z\sim6$ are imperative.

\subsection{SFR density at $z\sim6$}\label{subsec:sfrdevo}
As our observations are follow-up observations of UV-bright galaxies, we estimate the dust-obscured star formation activities of the UV-selected galaxies at $z\sim6$.
The focus of this paper, the contribution of dust-obscured star formation in the UV-selected galaxies, is complementary to the studies based on the IR-based source identification \citep[e.g.,][]{2020A&A...643A...8G,2021ApJ...909..165Z,2023arXiv230301658F} to comprehensively understand the dust-obscured star formation.
Our starting point is the galaxy UV luminosity function presented in \citet{2022ApJS..259...20H}.
The selection criteria adopted in their study allow the identification of not only blue $i$-dropout LBGs but also those with moderately red colors encompassing a $\beta_{\rm UV}$ range of $-3\leq\beta_{\rm UV}\leq1$.
We convert the UV luminosities of the galaxies into UV and IR luminosities in each UV absolute magnitude bin by utilizing both the $M_{\rm UV}$--$\beta_{\rm UV}$ relation including fainter galaxies in the literature (Figure \ref{fig:Muv_beta}) and IRX-$\beta_{\rm UV}$ relation (Figure \ref{fig:irx_beta}).
For the IRX-$\beta_{\rm UV}$ relation, we adopt the extinction curve reported in \citet{2020A&A...643A...4F} as a fiducial one due to its alignment with our results.
We estimate uncertainties using bootstrap techniques, ensuring the propagation of uncertainties in each relationship.
Subsequently, we derive the total (UV+IR) SFR from the UV and IR luminosities following the conversion factors in \citet{2014ARA&A..52..415M} with \citet{2003PASP..115..763C} IMF,
\begin{equation}
{\rm SFR}_{\rm UV} = 0.76\times10^{-28}L_{\rm UV} ({\rm erg}\,{\rm s}^{-1}\,{\rm Hz}^{-1})
\end{equation}
\begin{equation}
{\rm SFR}_{\rm IR} = 2.64\times10^{-44}L_{\rm IR} ({\rm erg}\,{\rm s}^{-1})
\end{equation}
and obtain the number densities of the galaxies as a function of their total SFR.
We then integrate the SFR function with a range from $M_{\rm UV}=-17$ mag following previous studies \citep[e.g.,][]{2012ApJ...754...83B,2021AJ....162...47B}.

\begin{figure}[t]
\begin{center}
\epsscale{1.15}
\includegraphics[width=8.5cm,bb=0 0 200 150, trim=0 1 0 0cm]{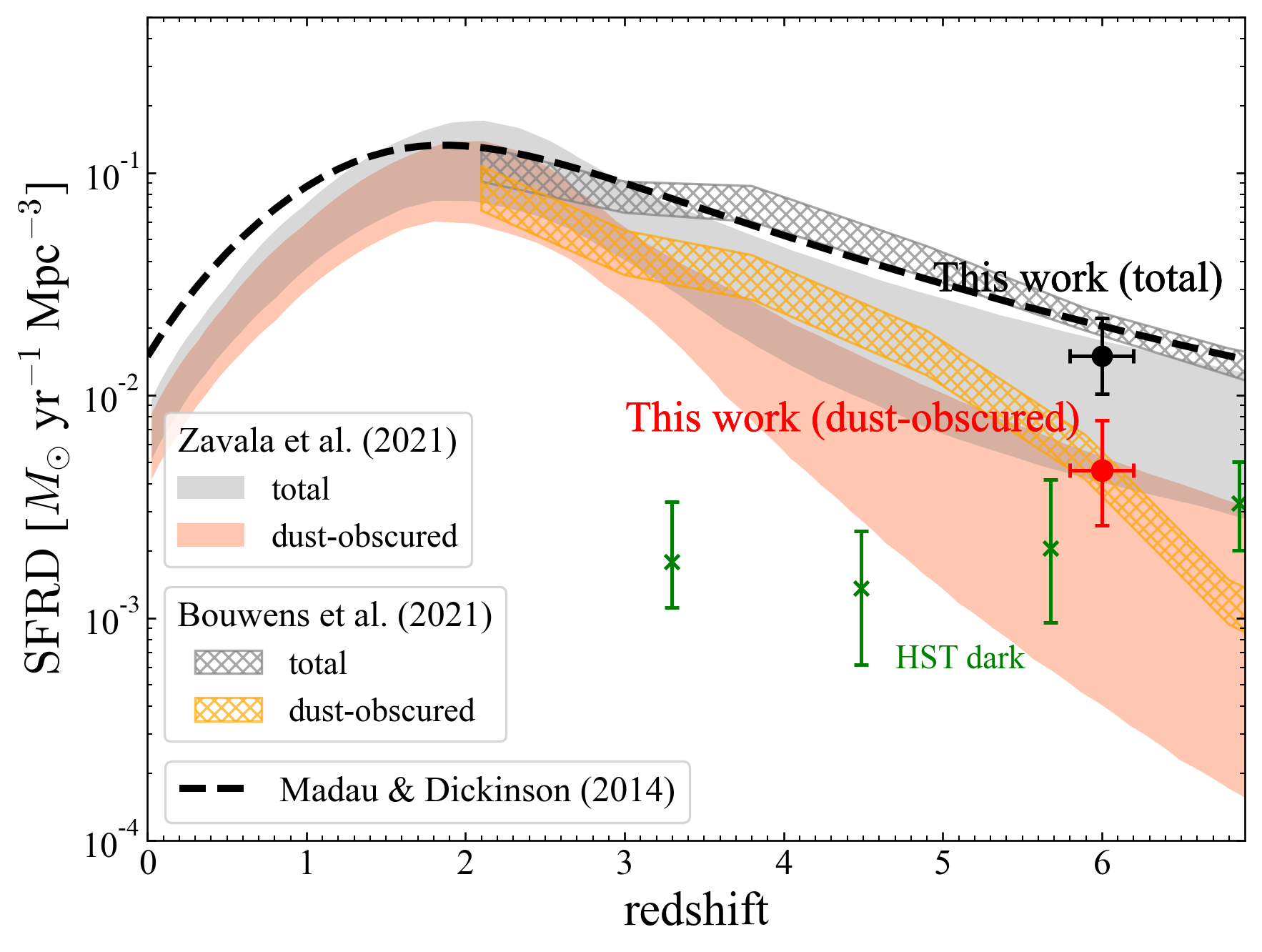}
\caption{Cosmic SFRD as a function of redshift.
The black and red circles indicate our results of the total and dust-obscured SFRDs, respectively, based on the IRX-$\beta_{\rm UV}$ relations consistent with our results and the $M_{\rm UV}$-$\beta_{\rm UV}$ relation constrained with our galaxies and the fainter galaxies in \citet{2014ApJ...793..115B}.
Previous studies are represented in the shaded areas for both the dust-obscured and unobscured SFRDs  \citep{2021AJ....162...47B,2021ApJ...909..165Z}.
The total SFRD presented in \citet{2014ARA&A..52..415M} is shown in the black dashed line.
The contribution of the {\it HST}-dark galaxies to the SFRD are also shown in the green crosses \citep{2023MNRAS.522..449B}.
Our results suggest that the dust-obscured star formation activity of LBGs does not significantly contribute to the total star formation rate, $\sim30\%$ at $z\sim6$.}
\label{fig:zSFRD}
\end{center}
\end{figure}

Figure \ref{fig:zSFRD} shows the SFRD as a function of redshift.
Our total and dust-obscured SFRD estimates are represented in the black and red-filled circles respectively.
Our analyses indicate that the dust-obscured star formation in UV-selected galaxies contributes $31_{-8}^{+8}\%$ to the total star formation at $z\sim6$.
This contribution is slightly larger than that predicted in \citet{2021AJ....162...47B}, $\sim24\%$.
The reason for this discrepancy comes from our adoption of the \citet{2020A&A...643A...4F} IRX-$\beta_{\rm UV}$ relation, which is shallower IRX but smaller intrinsic $\beta_{\rm UV}$ of -2.62 than the Meurer relation used in \citet{2021AJ....162...47B}.
Our estimation of the contribution of the dust-obscured star formation is near the upper bound of the results in \citet{2021ApJ...909..165Z} while their study focuses on dusty galaxies identified with ALMA. 

For a further comparison, we also calculate the SFRD with the different combinations of the $M_{\rm UV}$--$\beta_{\rm UV}$ and IRX-$\beta_{\rm UV}$ relation.
We test the following two cases with bluer UV slopes than our fiducial estimate above to consider the possibility that the slope is overestimated as discussed in Section \ref{subsec:IRXbeta}:
(1) linear $M_{\rm UV}$--$\beta_{\rm UV}$ and SMC IRX-$\beta_{\rm UV}$ relation (2) linear $M_{\rm UV}$--$\beta_{\rm UV}$ and Calzetti IRX-$\beta_{\rm UV}$ relation.
The calculated SFRDs are listed in Table \ref{tab:tab3}.
We find that the IRX-$\beta_{\rm UV}$ relation has a more profound impact on the resulting SFRD than the $M_{\rm UV}$--$\beta_{\rm UV}$ relation, but the maximum contribution of the dust-obscured star formation is $\sim45\%$ even with aggressive dust correction assumptions of (2), which is not preferred from the IRX measurements in this study.
Thus, we conclude that the contribution of the dust-obscured star formation activity from LBGs is not dominant and the assumption of the dust correction does not strongly change the total SFRD at $z\sim6$.
Note that our estimations of the dust-obscured star formation miss very dusty galaxies such as SMGs \citep[e.g.,][]{2020MNRAS.494.3828D} or {\it HST}-dark galaxies \citep[][]{2019Natur.572..211W} given our focus on the UV-bright galaxies.
Recent {\it JWST} investigation on {\it HST}-dark galaxies imply an almost constant contribution of the dust-obscured star formation activity from faint dusty galaxies from $z=4$ to $z=7$ \citep{2023MNRAS.522..449B}, although their contributions are predicted to be comparable with that of our estimation at $z\sim6$.

%
%
%
%
%
%

\subsection{Obscured fraction of the star formation}\label{subsec:fobs}
Apart from using the $M_{\rm UV}$-$\beta_{\rm UV}$ and IRX-$\beta_{\rm UV}$ relationships, the dust-obscured star formation in LBGs can be estimated through the direct compilation of the obscured fraction of the star formation, presented as a function of either stellar mass or absolute UV magnitude ($f_{\rm obs}$-$M_{\rm UV}$ or $f_{\rm obs}$-$M_{\ast}$). 
The obscured fraction is defined by the ratio of the obscured star formation rate (${\rm SFR}_{\rm IR}$) to the total (UV+IR) star formation rate (${\rm SFR}_{\rm UV+IR}$). 
The dust-obscured SFRD is directly derived via the UV luminosity function and $f_{\rm obs}$-$M_{\rm UV}$ relation in a similar way to that via the stellar mass function and $f_{\rm obs}$-$M_{\ast}$ relation utilized in some previous studies \citep[e.g.,][]{2023MNRAS.518.6142A}.

\begin{deluxetable}{cccc}
\tablecaption{Estimated SFRD under different assumptions \label{tab:tab3}}
\tablewidth{0pt}
\tablehead{
\multicolumn{2}{c}{assumption} & 
\colhead{total} &
\colhead{dust-obscured} 
}
\startdata
\multicolumn{2}{c}{$M_{\rm UV}$--$\beta_{\rm UV}$ \hspace{10pt}+\hspace{10pt} IRX-$\beta_{\rm UV}$} & \multicolumn{2}{c}{$[M_{\odot}\,{\rm yr}^{-1}\,{\rm Mpc}^{-3}]$} \\
\hline
B14+This work & \citet{2020AandA...643A...4F} & $-1.82_{-0.17}^{+0.17}$ & $-2.33_{-0.25}^{+0.23}$ \\
B14 extrapolation & SMC & $-1.72_{-0.17}^{+0.18}$ & $-2.06_{-0.27}^{+0.24}$ \\
B14 extrapolation & Meurer & $-1.73_{-0.22}^{+0.23}$ & $-2.13_{-0.50}^{+0.37}$ \\
\hline
 \multicolumn{2}{c}{$M_{\rm UV}$--$f_{\rm obs}$} & \multicolumn{2}{c}{$[M_{\odot}\,{\rm yr}^{-1}\,{\rm Mpc}^{-3}]$}\\
\hline
 \multicolumn{2}{c}{This work} & $-1.90_{-0.17}^{+0.19}$ & $-2.72_{-0.41}^{+0.40}$ \\ 
\enddata
\end{deluxetable}

To investigate the $f_{\rm obs}$-$M_{\rm UV}$ relation for the SERENADE galaxies, we split the sample into three distinct $M_{\rm UV}$ bins, $M_{\rm UV}$=[-24.5, -23.5], [-23.5, -22.0] and [-22.0, -20.5].
Owing to the limited observation in $88\,\mu$m within the $M_{\rm UV}$=[-22.0, -20.5] range, we conduct the stacking analysis for just the rest-frame 158$\mu{\rm m}$, rather than limiting the sample to the galaxies observed in both Band-6 and Band-8.
Following the methodologies in Section \ref{subsec:stacking} and \ref{subsec:dustflux}, we estimate stacked dust continuum fluxes and convert them to $L_{\rm IR}$ assuming an optically-thin MBB profile with $T_{\rm dust}=45\,{\rm K}$ and $\beta_{\rm dust}=1.8$ to perform a fair comparison with the previous results \citep{2020A&A...643A...4F,2023MNRAS.518.6142A}.
Then we calculate $f_{\rm obs}$ by translating the inferred $L_{\rm IR}$ into ${\rm SFR}_{\rm IR}$ and the average $L_{\rm UV}$ of the bin into ${\rm SFR}_{\rm UV}$.
We also calculate individual $f_{\rm obs}$ for reference.

\begin{figure}[t]
\begin{center}
\epsscale{1.15}
\includegraphics[width=8.5cm,bb=0 0 200 150, trim=0 1 0 0cm]{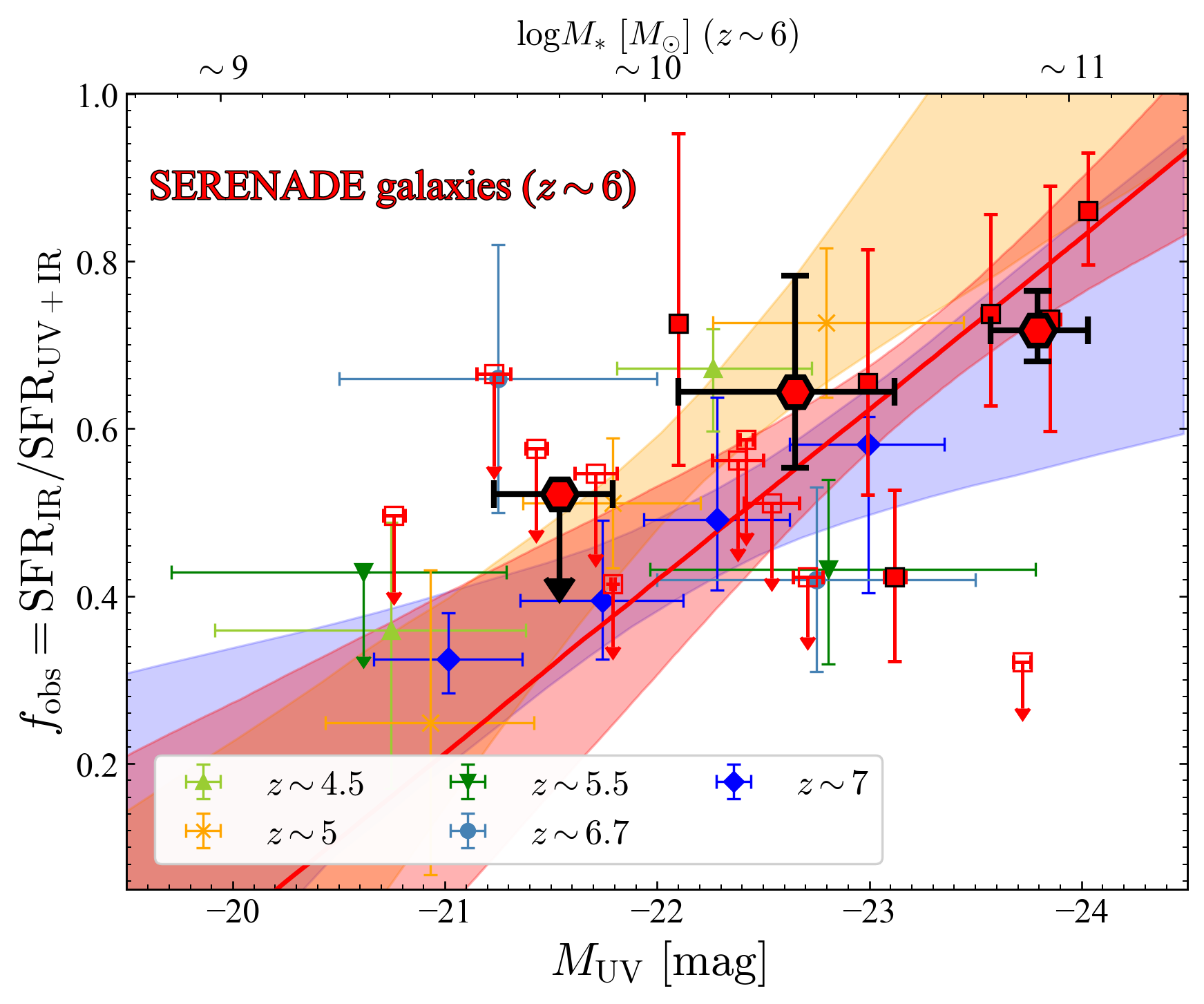}
\caption{Obscured fraction of the star formation against the absolute UV magnitude ($M_{\rm UV}$-$f_{\rm obs}$). 
Our individual measurements of the galaxies detected and undetected in the dust continuum are shown in the red-filled and open squares, respectively.
We also show the best fit and $1\sigma$ uncertainly of $M_{\rm UV}$--$f_{\rm obs}$ relation (red solid line and shaded region) derived from the stacked results (red pentagons).
We convert $M_{\ast}$ into $M_{\rm UV}$ using the scaling relation in \citet{2016ApJ...825....5S} for previous results for a direct comparison with our results. 
The stellar mass converted from $M_{\rm UV}$ at $z\sim6$ is also shown in the x-axis for reference.
The light green, orange, and green symbols are the results at $z\sim4.5$, 5, and 5.5. \citep[][Mitsuhashi et al. in prep]{2020A&A...643A...4F}. The blue and cyan symbols are the results at $z\sim6.7$ and $z\sim7$ from \citet{,2023arXiv230917386B} and \citet{2023MNRAS.518.6142A}, respectively.}
\label{fig:Muv_fobs}
\end{center}
\end{figure}

Figure \ref{fig:Muv_fobs} shows the derived individual and stacked $f_{\rm obs}$ values in red markers.
The best-fit $f_{\rm obs}$-$M_{\rm UV}$ relation and 1$\sigma$ uncertainties estimated by bootstrap are also shown in red solid line and shaded area.
Given that previous studies at $z\sim4.5$, 5.5, and 7 have measured the $f_{\rm obs}$ relative to the stellar mass, we convert them to $M_{\rm UV}$ based on the scaling relations in \citet{2016ApJ...825....5S}.
We note that the conversion is just for reference given the large uncertainties in the star formation histories of high-$z$ galaxies \citep[e.g.,][]{2022ApJ...927..170T,2023ApJ...957L..18S,2023arXiv230214155L}.
Our results show a good agreement in the range of $M_{\rm UV}\geq-23.1$ with the previous result at $z\sim5$, and add a constraint within the range of $M_{\rm UV}$=[-24.5, -23.5] corresponding to $M_{\ast}\sim10^{11}M_{\odot}$.
Our best-fit relation at $z\sim6$ shows a positive trend between the UV brightness and $f_{\rm obs}$.
This is similar to the previous results showing positive trends with $M_{\ast}-f_{\rm obs}$ at $z\sim4.5$, $z\sim5$, and $z\sim7$ from ALPINE, CRISTAL and REBELS survey \citep[][Mitsuhashi et al. in prep]{2020A&A...643A...4F,2023MNRAS.518.6142A}.
Our results show slightly higher $f_{\rm obs}$ than that of ALPINE galaxies at $z\sim5.5$ \citep{2020A&A...643A...4F} at the range of $\log M_{\ast}[M_{\odot}]=10$--11.1 corresponding $M_{\rm UV}=[-23.5, -22.0]$.
Selecting targets on the basis of Ly-$\alpha$ emission might induce a bias toward dust-poor galaxies, and result in the slightly lower $f_{\rm obs}$ in \citet{2020A&A...643A...4F}.
While a significant population (90\%) of the ALPINE galaxies within $\log M_{\ast}[M_{\odot}]=10$--11.1 are identified by Ly-$\alpha$ emission \citep{2020ApJS..247...61F}, only 50\% of the SERENADE galaxies show Ly-$\alpha$ emission.
\citet{2023arXiv230917386B} suggest the anti-correlation between $M_{\rm UV}$ and $f_{\rm obs}$ at $z\sim6.7$, which is contrary to the other results at $z>4$.
As described in Section \ref{subsec:Muv_beta}, the effect of scatter in the obscuration or geometry of their target galaxies may be one of the causes of this difference.

Based on the estimated $M_{\rm UV}$-$f_{\rm obs}$ relation and the UV luminosity function presented in \citet{2022ApJS..259...20H}, we calculate the total and the dust-obscured SFRD, following in Section \ref{subsec:sfrdevo}.
The calculated SFRDs are shown in Table \ref{tab:tab3}, corroborating the limited contribution of the dust-obscured star formation activity from LBGs ($\sim15\%$) as shown in Section \ref{subsec:sfrdevo}.

%
%
%
%
%
%
%
%
%
%
%
%

%
%
%
%
%
%
\begin{figure*}[htbp]
\begin{center}
\epsscale{1.15}
\includegraphics[width=16cm,bb=0 0 1000 650, trim=0 1 0 0cm]{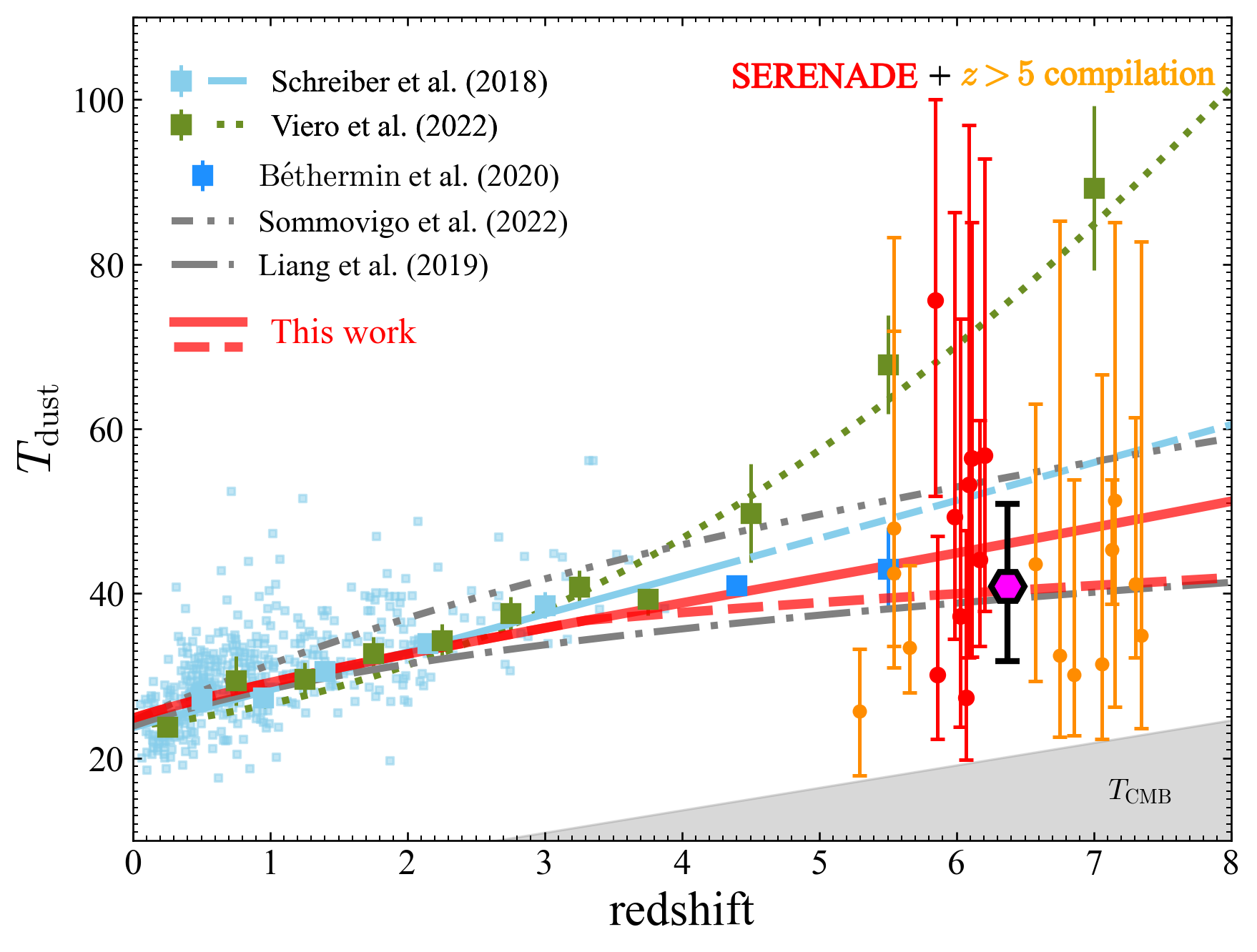}
\caption{Dust temperature as a function of the redshift. Individual results of the SERENADE galaxies are shown in the red circles and our measurement for $z\gtrsim5$ galaxies in the literature are shown in the orange markers \citep{2019PASJ...71...71H,2019MNRAS.487L..81L,2020MNRAS.498.4192F,2021ApJ...923....5S,2020MNRAS.493.4294B,2021MNRAS.508L..58B,2022MNRAS.515.1751W,2023arXiv230109659A}. The magenta pentagon shows the stacked results of the galaxies at $z\sim6$--7. For comparison, previous results based on the stacking analysis and the expected redshift evolution are shown in the blue, light blue, and green colors \citep{2018A&A...609A..30S,2020A&A...643A...2B,2022MNRAS.516L..30V}. Two model predictions of the redshift evolution are shown in the gray lines \citep{2019MNRAS.489.1397L,2022MNRAS.513.3122S}. The red lines represent the $T_{\rm dust}$ evolution calculated based on the analytical model of \citet{2022MNRAS.513.3122S} and observed redshift evolution of $t_{\rm gas}$ \citep{2020ARA&A..58..157T} and $Z$ \citep{2021ApJ...914...19S} under the assumption of continuous (solid) and no (dashed) $M_{\ast}-Z$ evolution at $z>3.3$.
The calculated $T_{\rm dust}$ evolution explains our result as well as the results at $z\lesssim5$.}
\label{fig:zTdust}
\end{center}
\end{figure*}

\subsection{$T_{\rm dust}$ evolution}\label{subsec:Tdevo}
In Figure \ref{fig:zTdust}, we plot $T_{\rm dust}$ as a function of the redshift. 
Among 28 galaxies covered in the multi-band observations, we plot 21 galaxies that are individually detected in Band-6 and/or Band-8, allowing us to measure $T_{\rm dust}$.
To estimate average $T_{\rm dust}$ at $z\sim6$--7, we also perform the stacking analysis of the galaxies observed in both 88 $\mu{\rm m}$ and 158 $\mu{\rm m}$.
From the 28 galaxies, we exclude four $z\sim5.5$ galaxies due to the different rest-frame wavelength coverage and two $z\sim6$ galaxies as in Section \ref{subsec:stacking}, resulting in a total of 22 galaxies for the stacking analysis.
We conduct an average stack and estimate $T_{\rm dust}$ and its uncertainty from MBB fitting with the Gaussian $\beta_{\rm dust}$ prior as described in Section \ref{subsec:mcmc}.
There is no major difference in the resulting $T_{\rm dust}$ when we use a median stacking instead of the average stacking.
Individual measurements from SERENADE (red circles) and additional $5<z<8$ galaxies from the literature (orange circles) are shown, along with the stacked result (magenta hexagon).
The stacked result suggests that an average dust temperature is $T_{\rm dust}=40.9_{-9.1}^{+10.0}\,{\rm K}$.
Our results support the linear relation of $\propto (1+z)^1$ \citep{2018A&A...609A..30S} rather than the quadratic relation $\propto (1+z)^2$ \citep{2022MNRAS.516L..30V}.
The individual measurements have a large scatter of $\Delta{T_{\rm dust}}\sim 12\,{\rm K}$ at $z=5$--8 \citep[see also,][]{2019PASJ...71...71H,2019MNRAS.487L..81L,2020MNRAS.498.4192F,2021ApJ...923....5S,2020MNRAS.493.4294B,2021MNRAS.508L..58B,2022MNRAS.515.1751W,2023arXiv230109659A}.

\section{Discussion}\label{sec:discussion}
As shown in the previous section, our result of $T_{\rm dust}\sim40\,{\rm K}$ supports a mild increase of $T_{\rm dust}$ as a function of redshift.
The theoretical predictions from \citet{2019MNRAS.489.1397L} also support our result, where they expect an increase of sSFR may result in an increasing trend of $T_{\rm dust}$ against the redshift.
Our result matches the evolution trend presented in \citet{2022MNRAS.513.3122S} with gas column density of $\log N_H[{\rm cm}^{-2}]\sim20$ under the optically thin assumption, where they predict that increasing trend of $T_{\rm dust}$ against redshift associates with higher SFR or shorter gas depletion timescale ($t_{\rm gas}$) at high-$z$ \citep[see also,][]{2024MNRAS.527...10V}.

To better understand the increasing trend of $T_{\rm dust}$, we calculate the $T_{\rm dust}$ evolution by combining the analytical model presented in \citet{2022MNRAS.513.3122S} with recent observational constraints.
Since an average stellar mass of $5<z<8$ galaxies are $M_{\ast}\sim10^{10}M_{\odot}$, we calculate $T_{\rm dust}$ evolution with $M_{\ast}=10^{10}M_{\odot}$.
We start with equations (7) and (9) in \citet{2022MNRAS.513.3122S},
\begin{equation}\label{eq7}
L_{\rm IR} = \left(\frac{M_{\rm dust}}{M_{\odot}}\right)\left(\frac{T_{\rm dust}}{8.5\,{\rm K}}\right)^{1/(4+\beta_{\rm dust})}
\end{equation}
and
\begin{equation}\label{eq6}
L_{\rm IR}=(1-e^{-\tau_{\rm eff}})L_{\rm 1500}^{\rm int}=(1-e^{-\tau_{\rm eff}})\mathcal{K}_{1500}{\rm SFR},
\end{equation}
where Equation (\ref{eq7}) can be obtained by integrating Equation (\ref{eq2}) with $\nu$.
The dust mass can be calculated from the gas mass, $M_{\rm dust}=DM_{\rm gas}$, where $D$ represents the dust-to-gas ratio proportional to the metallicity $D=D_{\odot}(Z/Z_{\odot})$ as shown in equation (3) in \citet{2022MNRAS.513.3122S}.
Here $\mathcal{K}_{1500}$ and $L_{\rm 1500}^{\rm int}$ are the conversion factor from the UV luminosity to the SFR (=${\rm SFR}/L_{1500}$) and the intrinsic UV luminosity, respectively.
$\tau_{\rm eff}$ and $D_{\odot}$ represent the effective dust attenuation optical depth at 1500\AA\ and Galactic dust-to-gas ratio, respectively.
A combination of these equations results in 
\begin{equation}\label{eq5}
T_{\rm dust} = 8.5\left[\left(1-e^{-\tau_{\rm eff}}\right)\frac{\mathcal{K}_{1500}}{D_{\odot}}\left(\frac{t_{\rm gas}}{{\rm yr}^{-1}}\cdot\frac{Z}{Z_{\odot}}\right)^{-1}\right]^{1/(4+\beta_{\rm dust})}
\end{equation}
where $t_{\rm gas}$ represents the gas depletion timescale (=$M_{\rm gas}/{\rm SFR}$).
For the redshift evolution of the metallicity, we assume a power-law dependence of $d\log({\rm O}/{\rm H})/dz=-0.11$ with a $z=0$ value of $12+\log({\rm O}/{\rm H})=8.77$ at $M_{\ast}=10^{10}M_{\odot}$ \citep{2021ApJ...914...19S}, which is consistent with recent {\it JWST} observations \citep{2023arXiv230112825N,2023arXiv230408516C}.
As the metallicity dependence with redshift is still uncertain at $z>5$ \citep{2023arXiv230112825N,2023arXiv230408516C}, we also compute the redshift evolution under the assumption of no metallicity evolution at $z>3.3$.
We use $\mathcal{K}_{1500}=1.4\times10^{10}\,L_{\odot}/(M_{\odot}\,{\rm yr}^{-1})$ following in \citet{2020ARA&A..58..157T} for the Chabrier IMF.
Assuming $\beta_{\rm dust}=1.8$ and $D_{\odot}=1/162$ \citep{2014A&A...563A..31R}, equation (\ref{eq5}) becomes
\begin{equation}
T_{\rm dust} = 19.55\times\left[10^{0.11z}(1+z)^{0.98}(1-e^{-\tau_{\rm eff}})\right]^{1/5.8}
\end{equation}
Here we use $\tau_{\rm eff}=1.3$ to match the $T_{\rm dust}$ observation at $z=0.5$, 27.2\,K, in \citet{2018A&A...609A..30S} for galaxies with $M_{\ast}=10^{9.5}$--$10^{10}M_{\odot}$. 
The $\tau_{\rm eff}$ value can be converted to ${\rm IRX}=0.56$ with
\begin{equation}
{\rm IRX} = L_{\rm IR}/L_{\rm 1500}^{\rm obs} = (1-e^{-\tau_{\rm eff}})/e^{-\tau_{\rm eff}}
\end{equation}
based on equation (\ref{eq6}), which is 
consistent with the measurements up to $z\sim6$ \citep{2014MNRAS.437.1268H,2017ApJ...850..208W,2020A&A...643A...4F,2023MNRAS.518.6142A}.
We plot the calculated $T_{\rm dust}$ evolution in Figure \ref{fig:zTdust}.
The calculated $T_{\rm dust}$ matches our observational estimate for $z\sim6$--7 galaxies ($T_{\rm dust}=40.9_{-9.1}^{+10.0}\,{\rm K}$).

The observed scatter of $T_{\rm dust}$ among $z\sim6$--7 galaxies ($\Delta T_{\rm dust}=12\,{\rm K}$) can also be naturally explained by dispersions in the $M_{\ast}-Z$ and $M_{\ast}-{\rm SFR}$  (so-called main sequence) relations.
The dispersion from the main sequence (MS) of $\Delta{\rm MS}=\pm0.6\,{\rm dex}$ \citep[see][]{2020ARA&A..58..157T} results in the change of $\Delta T_{\rm dust}\sim\pm6\,{\rm K}$.
A variance of $\Delta Z=\pm0.3\,{\rm dex}$ \citep[see][]{2021ApJ...914...19S} from an average $M_{\ast}-Z$ relationship also changes $T_{\rm dust}$ by $\Delta T_{\rm dust}\sim\pm6\,{\rm K}$.
Based on the fundamental mass-metallicity relation, the metallicity decreases with increasing SFR at the fixed stellar mass \citep[e.g.,][]{2013ApJ...765..140A}.
Therefore, the positive offset from the MS may correlate with the negative offset in the average $M_{\ast}-Z$ relation, and both offsets result in the positive offset in $T_{\rm dust}$.
A sum of the dispersions in the MS and $M_{\ast}-Z$ relation is $\Delta T_{\rm dust}=\pm12\,{\rm K}$, which is comparable to our observational result for the scatter in $T_{\rm dust}$.
We note that the assumption of $\beta_{\rm dust}$ (the Gaussian prior with $\langle\beta_{\rm dust}\rangle=1.8$ and $\sigma_{\beta_{\rm dust}}=0.5$) does not strongly contribute to the spread of $T_{\rm dust}$ given a small dispersion in the best-fit $\beta_{\rm dust}$ values of $\sim0.1$ corresponding to $\Delta T_{\rm dust}\sim3\,{\rm K}$.
The sampling of the continuum SED at longer wavelengths is necessary to obtain more precise $\beta_{\rm dust}$ values.
Interestingly, the high $T_{\rm dust}$ in the $z=8.3$ MACS0416-Y1 \citep[][]{2019ApJ...874...27T,2020MNRAS.493.4294B} can also be explained by our calculation, since its high sSFR (${\rm sSFR}\sim2.0\times10^{-7}\,{\rm Gyr}^{-1}$, corresponding to $\Delta {\rm MS}\sim+1.5$--1.6 dex; \citealt{2014ApJS..214...15S,2022MNRAS.516..975T}) and low metallicity ($Z\sim0.02$--$0.2Z_{\odot}$) are consistent with the measured lower limit of the dust temperature of $T_{\rm dust}>80\,{\rm K}$.
Although our targeted galaxies do not have stellar mass $M_{\ast}$ or the metallicity $Z$ estimates due to the lack of rest-frame optical coverage, future {\it JWST} observations may be able to confirm what changes $T_{\rm dust}$.

%
%
%
%
%
%
\section{Summary and Conclusions}\label{sec:summary}
In this paper, we have examined the dust continuum emissions of 28 LBGs at $z\sim5$--8 by utilizing multi-band dust-continuum observations obtained with ALMA.
We conduct the MBB fitting with the MCMC algorithm to constrain parameters characterizing the FIR SEDs such as $T_{\rm dust}$ and $M_{\rm dust}$.
The 11 galaxies newly observed in our program, SERENADE, are originally UV-selected but cover a wide range of $L_{\rm IR}$ that spans from dust-poor galaxies (non-detection) to HyLIRG-class galaxies.
The brightest galaxies in $L_{\rm IR}$ among our target galaxies exhibit intense starbursts comparable with the dusty star-forming galaxies at $z\sim7$.
From the individual measurements as well as the stacking analysis, we have found:

\begin{itemize}
\item The IRX-$\beta_{\rm UV}$ relation derived from multi-band observations at $z\sim6$ appears to be $\sim1\,{\rm dex}$ lower than that at $z\sim3$ at $\beta_{\rm UV}\sim0$.
Our observational constraints show good agreement with the previous $z\sim5.5$ results based on single-band ALMA observations. The dust-obscured star formation calculated with the $M_{\rm UV}$-$\beta_{\rm UV}$ relation including fainter galaxies in literature and the IRX-$\beta_{\rm UV}$ relations consistent with our results suggest $31_{-8}^{+8}\%$ contribution to the total SFRD at $z\sim6$.

\item The $M_{\rm UV}$-$f_{\rm obs}$ relation at $z\sim6$ is consistent with the results at $z\sim5$ and $z\sim7$.
The contribution of dust-obscured star formation among LBGs based on the $M_{\rm UV}$-$f_{\rm obs}$ relation comprises $15_{-8}^{+17}\%$ of total SFRD, consistent with our estimates based on the $M_{\rm UV}$-$\beta_{\rm UV}$ and IRX-$\beta_{\rm UV}$ relations within errors.

\item We present dust temperature measurements of LBGs at $z=5-8$. Our $T_{\rm dust}$ measurements remain consistent with a linear redshift evolution [$\propto(1+z)^{1}$].
On the basis of the analytical model presented in \citet{2022MNRAS.513.3122S}, we calculate the average evolution of $T_{\rm dust}$ by combining $t_{\rm gas}$ \citep{2020ARA&A..58..157T} and $Z$ evolution \citep{2021ApJ...914...19S} for galaxies with $M_{\ast}\sim10^{10}M_{\odot}$ consistent with the recent {\it JWST} measurements in \citet{2023arXiv230112825N} and \citet{2023arXiv230408516C}.
The calculated $T_{\rm dust}$ evolution exhibits an excellent agreement with our measurements at $z>5$ as well as literature results at $z<5$.
The large observed scatter in $T_{\rm dust}$ can be interpreted as the results from the scatter around the star formation main sequence and the average mass-metallicity relation.

\end{itemize}

Since ALMA has capabilities of high-frequency observation in Band-9/10, future ALMA high-frequency observations for a large sample of galaxies will further improve $T_{\rm dust}$ constraints and enable us to reach more precise conclusions.
The recent {\it JWST} discovery of a large fraction of the broad line AGN \citep[Type-1,][]{2023arXiv230311946H,2023arXiv230801230M,2023arXiv230605448M,2023ApJ...954L...4K} may suggest that the IR luminosity and the dust temperature are boosted by the possible presence of Type-2 AGN, as shown in \citet[][see also, \citealp{2023MNRAS.523.4654T,2023arXiv230915150T}]{2021ApJ...921...55M}, although the estimated luminosity-weighted dust temperatures in this work have a very good agreement with those without strong effect of the AGN in the cosmological simulation \citep{2021MNRAS.503.2349D}.
Spetially-resolved observations in Band-9/10 may allow us to evaluate the AGN contribution to the radiation in FIR wavelength.
Also, future ALMA wideband sensitivity upgrades will enable us to access the very faint dust continuum from high-$z$ galaxies to constrain the IRX or $f_{\rm obs}$ values of UV faint galaxies. 

\acknowledgments
We thank L. Sommovigo and K. Kohno for giving us helpful comments. This paper makes use of the following ALMA data: ADS/JAO.ALMA\#2015.1.00540.S, \#2015.1.01406.S, \#2016.1.00954.S, \#2017.1.00190.S, \#2017.1.00508.S, \#2017.1.00697.S, \#2017.1.00775.S, \#2019.1.01634.L, \#2021.1.01297.S, \#2021.1.00318.S, \#2022.1.00522.S. 
ALMA is a partnership of ESO (representing its member states), NSF (USA), and NINS (Japan), together with NRC (Canada), MOST and ASIAA (Taiwan), and KASI (Republic of Korea), in cooperation with the Republic of Chile. The Joint ALMA Observatory is operated by ESO, AUI/NRAO, and NAOJ.
IM is financially supported by Grants-in-Aid for Japan Society for the Promotion of Science (JSPS) Fellows
(KAKENHI Number 22KJ0821).
YH is supported by JSPS KAKENHI Grant Number 21K13953.
H.U. acknowledges support from the JSPS KAKENHI grant (20H01953, 22KK0231).
ANID - Millennium Science Initiative Program - ICN12\_009 (FEB), CATA-BASAL - FB210003 (FEB), and FONDECYT Regular - 1200495 (FEB).
Data analysis was in part carried out on the Multi-wavelength Data Analysis System operated by the Astronomy Data Center (ADC), National Astronomical Observatory of Japan.

\appendix
\renewcommand{\thesection}{A}

\section{flux comparison}\label{appendix:fluxcomp}
Because the amplitude of the visibilities at the zero baseline length could be a good indicator of the total flux, we also measure the fluxes from visibility data with {\sc UVMultifit} assuming S{\'e}rsic index $n=1$ just to check the flux recovery.
Details of this visibility-based analysis will be presented in Mitsuhashi et al. (in prep).

\begin{figure*}[htbp]
\begin{center}
\epsscale{1.15}
\includegraphics[width=16cm,bb=0 0 1000 650, trim=0 1 0 0cm]{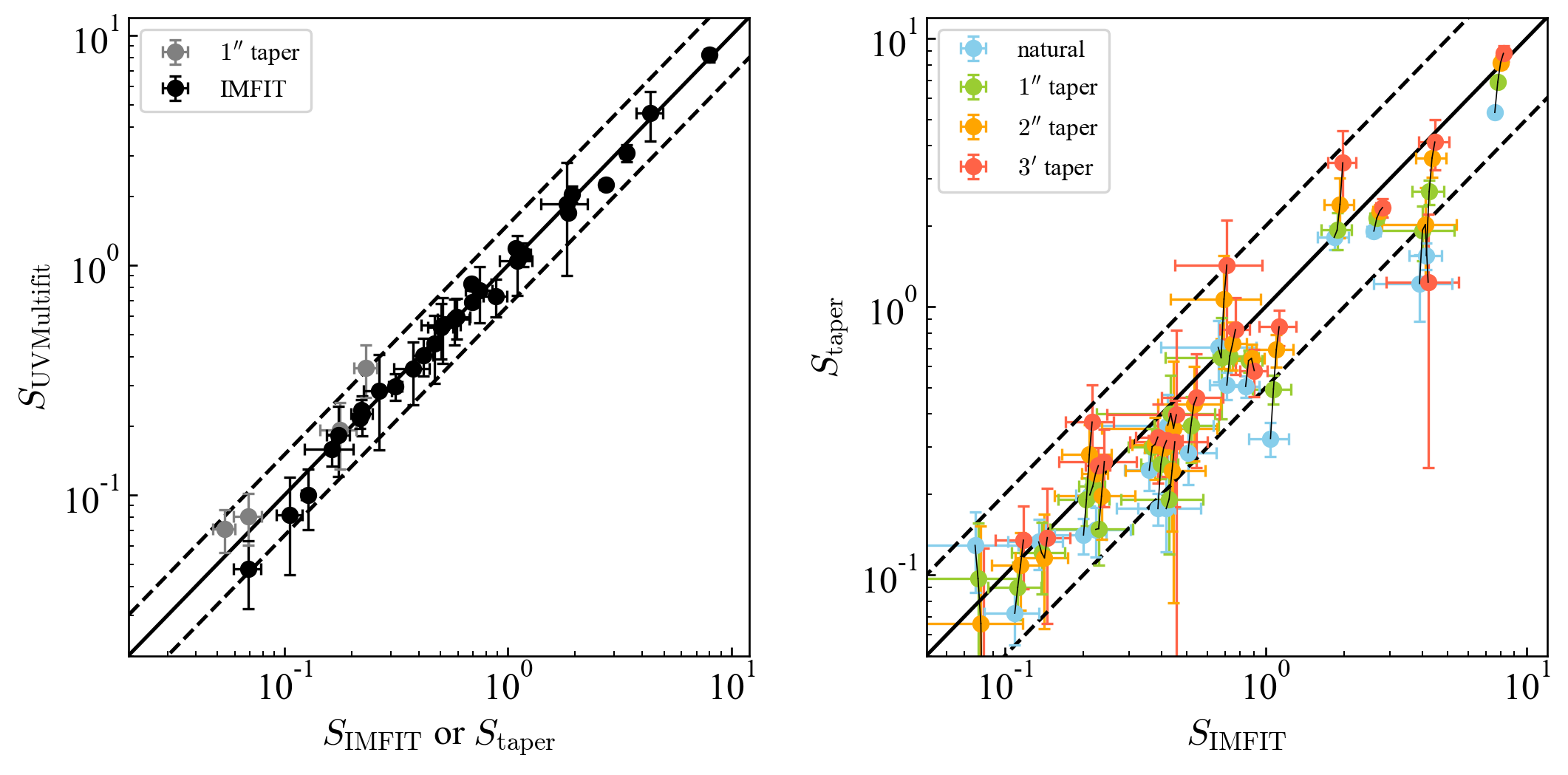}
\caption{Flux comparison between the different methodologies. (left) Fluxes computed with visibility fitting vs those with image-based {\sc imfit} or peak fluxes. One-to-one relation and $\pm0.3\,{\rm dex}$ are shown in the black solid and dashed lines. (right) Fluxes computed from {\sc imfit} and peak fluxes in the different taper scales of $0''$ (natural weighting), $1''$, $2''$, and $3''$. We again show one-to-one relation and $\pm0.5\,{\rm dex}$ in black solid and dashed lines.}
\label{fig:fluxcomp}
\end{center}
\end{figure*}

In Figure \ref{fig:fluxcomp}, we compare the flux densities calculated with the different methodologies.
Figure \ref{fig:fluxcomp} (left) compares the fluxes derived from visibility fitting with {\sc UVMultifit} and from the image-based measurement.
As described in Section \ref{sec:analysis}, we apply two different ways to measure fluxes in the image, the {\sc CASA} task {\sc imfit} or peak flux with the Tapered images. 
We find a very good consistency between visibility- and image-based flux densities we used in this work.

Figure \ref{fig:fluxcomp} (right) plots the flux densities based on the {\sc CASA/imfit} against the peak flux in the images with different taper scales.
The taper scale needed to recover total fluxes changes depending on the spatial extent of the sources.
The uncertainties in the taper scales of $>2''$ tend to be larger than those from {\sc imfit} although taper scales of $1''$--$3''$ are necessary to recover the fluxes.

%
%
%
%
%
%
%
%
%
%
%
%


\clearpage
\bibliography{ref.bib}{}
\bibliographystyle{apj.bst}

\end{document}